\documentclass[%
 reprint,
 amsmath,amssymb,
 aps,
 nofootinbib
]{revtex4-2}
\usepackage{float}
\usepackage{graphicx}% Include figure files
\usepackage{dcolumn}% Align table columns on decimal point
\usepackage{bm}% bold math
\usepackage{soul} % to cross a text; command \st{....}
\usepackage{color}% print color text
\definecolor{vs}{rgb}{0.1,0.4,0.1}                  % dark green
\newcommand{\del}[1]{}                             % to accept all deletions comment \newcommand{\del}[1]{\textcolor{vs}{\st{#1}}} with % and uncomment this line
\numberwithin{equation}{section}
\renewcommand\theequation{\arabic{section}.\arabic{equation}}
\usepackage{hyperref}% add hypertext capabilities
\usepackage{verbatim}
\newcommand\wordcount{\verbatiminput{\jobname.sum}}
\begin{document}
%\preprint{}
\begin{minipage}[h]{\textwidth}
  Proc. R. Soc. A {\bf 479}: 20230366 (2023). \\
  \protect\href{https://doi.org/10.1098/rspa.2023.0366}{https://doi.org/10.1098/rspa.2023.0366}
  \mbox{}\\
\end{minipage}
\title{Exact Solutions to Fall of Particle to Singular Potential: Classical vs. Quantum Cases}


\author{Michael I. Tribelsky}
\email[]{\mbox{E-mail: mitribel\_at\_gmail.com \@(replace ``\_at\_" by @)}
}
\homepage[]{\\https://polly.phys.msu.ru/en/labs/Tribelsky/}
%\thanks{}
\affiliation{%Faculty of Physics,
M. V. Lomonosov Moscow State University, Moscow, 119991, Russia}
\affiliation{Center for Photonics and 2D Materials, Moscow Institute of Physics and Technology, Dolgoprudny 141700, Russia}
%\affiliation{National Research Nuclear University MEPhI (Moscow Engineering Physics Institute), Moscow, 115409, Russia}
%\affiliation{Landau Institute for Theoretical Physics RAS, Chernogolovka, Moscow Region 142432, Russia}

\date{\today}

\begin{abstract}{Exact solutions describing a fall of a particle to the center of a non-regularized singular potential in classical and quantum cases are obtained and compared. We inspect the quantum problem with the help of the conventional Schr\"{o}dinger's equation. During the fall, the wave function spatial localization area contracts into a single zero-dimensional point. For the fall-admitting potentials, the Hamiltonian is non-Hermitian. Because of that, the wave function norm occurs time-dependent. It demands an extension to this case of the continuity equation and rules for mean value calculations. Surprisingly, the quantum and classical solutions exhibit striking similarities. In particular, both are self-similar at the particle energy equals zero. The characteristic spatial scales of the quantum and classical self-similar solutions obey the same temporal dependence. We present arguments indicating that these self-similar solutions are attractors to a broader class of solutions, describing the fall at finite energy of the particle. %In addition to the collapsing solutions, exact expanding solutions corresponding to the particle's escape from the singularity exist for both problems. The collapse and escape are continuously linked to each other: the completion of one process marks the beginning of the other.
}
\end{abstract}
%
%We obtain, inspect, and compare exact solutions describing the fall of a particle into the center of a singular potential (collapse) in classical and quantum cases. Surprisingly, the quantum and classical solutions exhibit striking similarities. In particular, self-similar properties are observed at zero energy of the particle, and their characteristic spatial scales exhibit the same temporal dependence. In addition to the collapsing solutions, exact expanding solutions corresponding to the particle's escape from the singularity exist for both problems. We show that the collapse and escape processes are continuously linked to each other: the completion of one process marks the beginning of the other. In the quantum collapse, wave functions of the exact solutions have time-dependent norms that demand generalization of the conventional continuity equations and rules for mean value calculations.

%\pacs{}
%\keywords{}

\maketitle
\section{Introduction}

Collapses, i.e. spatio-temporal evolutions of smooth solutions resulting in the formation of singularities, happen in various physical phenomena and play a significant role there. Suffice to mention implosion of spherical and cylindrical shock waves~\cite{guderley1942starke,stanyukovich,zeld_raizer}; the collapse of
bubbles in a liquid~\cite{zeld_raizer,rayleigh1917_buble,hunter1960collapse,Zababakh_buble}; the self-focusing in nonlinear optics~\cite{askaryan1962cerenkov,Chiao1964,Kelley1965}; gradient catastrophe of acoustic waves~\cite{arnold2013mathematical}; and Langmuir's collapse in plasma physics~\cite{zakharov1972collapse}. For more examples and their discussions, see, e.g. review~\cite{zakharov2012_Collapse_UFN} and references therein.

Among various collapses, a fall of a quantum particle to the center of a singular spherically symmetric potential, also known as {\it quantum collapse}, has a special significance: the potentials admitting the quantum collapse make the Hamiltonian non-Hermitian; see below, Sec.~\ref{sec:quant_coll}. {As a result, Schr\"{o}dinger's equation with this potential fails to produce the ground state~\cite{landau2013quantum,book:Morse_Feshbach}. %The effect is a manifestation of {\it the quantum anomaly}: violation of the scale invariance of the classical Hamiltonian at the problem quantization~\cite{PhysRevD.48.5940,PhysRevLett.85.1590}. 
This gives rise to very unusual properties of wave functions so that the conventional rules for the mean value calculations and continuity equation cannot be used in this case and demand reconsideration; see below.} %Thus, a study of the quantum collapse should shed new light on these issues and give a more profound understanding.

{In 2023, revising fundamental concepts of quantum mechanics might seem peculiar, to say the least. Therefore, we want to clarify that the suggested changes have no bearing on the common problems of non-relativistic quantum mechanics. They only affect the particular solutions to Schr\"{o}dinger's equation with the non-Hermitian Hamiltonian to overcome the intrinsic contradictions of the conventional rules arising in this case. }

The quantum collapse is a rare but not the only case of non-Hermitian Hamiltonians in quantum mechanics. They may also be introduced in other essentially time-dependent problems, e.g. in the $\alpha$-decay, where a complex value of energy corresponds to the decaying in time probability to find the particle in a quasi-discrete level~\cite{Betan2012} or, in more general terms, in various manifestations of resonant scattering of particles by potentials with quasi-discrete levels, known as Fano resonances~\cite{Fano1961}, etc. 

In contrast to conventional Hermitian Hamiltonians, each case of Schr\"{o}dinger's equation with a non-Hermitian Hamiltonian requires an individual consideration valid, generally speaking, only for a given problem. Accordingly, the approach of the present paper is explicitly developed for the fall to the center. However, the method employed here is much broader and may also be applied to other problems, whose examples are mentioned above.

Study of Schr\"{o}dinger's equation with collapse-admitting potentials has a long-lasting history~\mbox{\cite{shortley1931inverse,case1950singular,guggenheim1966inverse,levy1967electron,Efimov1971,alliluev1972problem,parisi1973anomalous,schwartz1976almost,van1978bound,wu1994inverse,song2000unitary,PhysRevLett.85.1590,dong2005algebraic,AvilaAoki2009,astrakharchik2015quantum,Malomed2018,sakaguchi2011suppression2,sakaguchi2013suppression3,sakaguchi2011suppression,shamriz2020suppression,Shamriz2020}.} Its results are discussed in reviews~\cite{frank1971singular,zakharov2012_Collapse_UFN} and enter text-books~\cite{book:Morse_Feshbach,landau2013quantum}.
Nonetheless, most of them are based on various regularization procedures (cutoff of a singular potential at the vicinity of the singularity, a shift of the boundary conditions from the singular point to its proximity, incorporation of nonlinear terms, etc.). On the other hand, any regularization procedure implies (explicitly or implicitly) that, at vanishing regularization parameters, the regularized solutions converge to non-regularized ones. This is not the case for the quantum collapse: at the point of the potential singularity, the wave functions do not have any definite limit~\cite{book:Morse_Feshbach,landau2013quantum}. Therefore, a continuous transition from a regularized problem to its original non-regularized version becomes impossible. Thus, the fundamental question of whether a spatial localization area for a wave function obeying Schr\"{o}dinger's equation indeed can collapse to a zero-dimensional point remains open. %{Moreover, the nonexistence of the limit for the wave functions at the singularity, seemingly, contradicts to the possibility of the localization area to collapse to the same singularity (we will resolve this contradiction; see the last paragraph of Sec.~\ref{sec:quant_coll}).}

The argument that, close to the singularity, the collapse conditions usually are violated does not compromise the problem. The issue is common to all collapses. Its resolution is well known: the collapse-admitting problem describes the most physically significant part of the dynamics, lasting as long as the spatial scale of the solution remains larger than the one where the collapse-breaking terms become essential. {At the same time, the problem description in terms of the collapse-admitting approach is more simple, informative, and convenient, than those based on the incorporation of the collapse-breaking corrections~\cite{zakharov2012_Collapse_UFN}.}

{Therefore, here we {\it intentionally\/} avoid any regularization. The questions we answer in this study are whether there are any {\it exact\/} collapse-exhibiting solutions to Schr\"{o}dinger's equation valid in all space and, if so, whether they agree with the fundamentals of quantum mechanics, despite the non-Hermitian Hamiltonian.}

{We obtain a set of such solutions. However, they are too unusual to allow straightforward interpretations. The main goal of the publication is to draw the community’s attention to the {\it interpretations} of the solutions rather than the solutions themselves. We did our best to interpret them, but we do not claim the given interpretation is ultimate...}

The paper has the following structure. In Sec. \ref{sec:class_coll}, we discuss the classical problem. In Sec.~\ref{sec:rad_loss}, we inspect a specific example of this case: a fall of an electron to a heavy dipole and reveal the role of radiative losses. In Sec.~\ref{sec:quant_coll}, we formulate the quantum problem. In Sec.~\ref{sec:quasi-class}, we derive the applicability condition for applying the quasi-classical approximation to the problem in question. In Sec.~\ref{sec:wave_function}, we obtain a family of, exhibiting collapse, exact solutions to Schr\"{o}dinger's equation. In Sec.~\ref{sec:norm}, we inspect the associated with these solutions peculiarities of time-dependent norms. Sec.~\ref{sec:mean_values} is devoted to calculations of mean values with the help of wave functions with time-dependent norms. In Sec.~\ref{sec:contin}, we describe and discuss the modification of the continuity equation required in this case. In Sec.~\ref{sec:examp}, we preset and inspect specific examples of the solutions to the quantum problem. Sec.~\ref{sec:concl} is devoted to conclusions. Appendix contains some ancillary calculations.

\section{Classical problem\label{sec:class_coll}}

To begin with, we consider classical collapse. Let us recall its main features~\cite{landau1976mechanics}. The spherical symmetry of the problem results in the conservation of the particle angular momentum $\mathbf{M} = \mathbf{r}\times\mathbf{p}$. It means that $\mathbf{r}(t)$ is always perpendicular to the constant vector $\mathbf{M}$, i.e. the particle trajectory is a two-dimensional curve. Then, it is convenient to describe this motion in a polar coordinate system with the origin at the center of the potential $U(r)$. Next, the angular momentum conservation leads to the following relation:
\begin{equation}\label{eq:dphi/dt}
\dot{\varphi}=\frac{M}{m r^2},
\end{equation}
{where dot stands for $d/dt$, and $m$ is the particle mass.}

{The energy integral has the form:
\begin{equation}\label{eq:E}
  E = \frac{m}{2}\left(\dot{r}^2+r^2\dot{\varphi}^2\right) + U(r) \equiv \frac{m\dot{r}^2}{2} + U_{\rm eff}(r).
\end{equation}
Here
\begin{equation}\label{eq:Ueff}
  U_{\rm eff}(r) = U(r)+\frac{m r^2\dot{\varphi}^2}{2} \equiv  U(r) + \frac{M^2}{2mr^2},
\end{equation}
see Eq.~\eqref{eq:dphi/dt}.}

{The introduced  $U_{\rm eff}(r)$  makes it possible to exclude $\varphi$ from the energy integral.} The obtained equation is one-dimensional and readily integrated for any $U(r)$. The result is as follows:
\begin{eqnarray}
   t-t_{\rm ini} &=& \pm \int_{r_{\rm ini}}^r\frac{dr'}{\sqrt{\frac{2}{m}\left[E-U_{\rm eff}(r')\right]}}\label{eq:t(r)_class}\\
   & \equiv & \pm \int_{r_{\rm ini}}^r\frac{dr'}{\sqrt{\frac{2}{m}\left[E-U(r')\right]-\frac{M^2}{m^2r'^2}}}; \nonumber
\end{eqnarray}
where $E$ stands for the particle energy, {and \mbox{$r_{\rm ini} \equiv r(t_{\rm ini})$} is the initial condition.} Note that we keep two signs of the square root in Eq.~\eqref{eq:t(r)_class}. {It means that the solution Eq.~\eqref{eq:t(r)_class} has two branches, so that both $t-t_{\rm ini}$ and $dr/dt$ may have any sign.}

{This fact is essential. To understand that, consider a finite motion of the particle corresponding to nonlinear oscillations of $r$ between $r_{\rm min}$ and $r_{\rm max}$, where $r_{\rm min,\,max}$ are two sequential roots of the equation \mbox{$E=U_{\rm eff}(r)$}, and $r_{\rm min}<r_{\rm max}$. The following expression gives the period of the oscillations:
\begin{equation}\label{eq:period_class}
  T = 2\int_{r_{\rm min}}^{r_{\rm max}}\frac{dr}{\sqrt{\frac{2}{m}\left[E-U_{\rm eff}(r)\right]}}.
\end{equation}}

{When $r(t)$ increases, $dr/dt>0$. It corresponds to the plus sign in Eq.~\eqref{eq:t(r)_class}. In contrast, the stage of the motion when $r(t)$ decreases corresponds to the minus sign. The conclusion is that at the return points, namely at \mbox{$r=r_{\min}$} and \mbox{$r=r_{\rm max}$,} a transition between the solution branches takes place. Since for both branches $dr/dt = 0$ at $r=r_{\rm min,\,max}$, see Eq.~\eqref{eq:t(r)_class}, %at finite $r=r_{\rm min,\,max}$,
the transition does not violate smoothness of the dependence $r(t)$. Thus, every time, when $r$ reaches a given value with the same sign of $dr/dt$, $t_{\rm ini}$ in Eq.~\eqref{eq:t(r)_class} increases by the period of oscillations $T$. Bearing this in mind, we proceed with the discussion of the collapse.}

The expression under the square root in Eq.~\eqref{eq:t(r)_class} must be non-negative. Then, the point $r=0$ is accessible, i.e. collapse is possible, provided
\begin{equation}\label{eq:class_cond}
  [r^2U(r)]_{r\rightarrow 0} \leq -{M^2}/(2m).
\end{equation}
{For definiteness,} in what follows, we consider the potential
\begin{equation}\label{eq:U}
  U(r) = -\beta/r^2;\;\; (\beta > 0),
\end{equation}
Notably, that potential Eq.~\eqref{eq:U} does exist in nature. For example, it describes the interaction of a point electric charge with a particle possessing zero total electric charge but a finite dipole moment~\cite{sakaguchi2011suppression}; see below, Sec~\ref{sec:rad_loss} {(the issue is essential in understanding electron capture by a polar molecule~\cite{levy1967electron}). It also arises in many other physical problems~\cite{PhysRevD.68.125013}, such as certain quantum three-body problems~\cite{Efimov1971,efimov1973energy}, the physics of cold atoms~\cite{denschlag1997scattering,kulkarni2011cold}, polymer physics~\cite{nisoli2014attractive}, the near-horizon problem for certain black holes~\cite{claus1998black}, etc. Thus, the collapse in this potential has practical importance in various branches of physics.}

The application of condition Eq.~\eqref{eq:class_cond} to potential Eq.~\eqref{eq:U} gives rise to the following {inequality:} $ \beta \geq \frac{M^2}{2m}$. {Note that, at $\beta < \frac{M^2}{2m}$, $U_{\rm eff}(r)>0$, i.e. the effective potential becomes repulsive. Only a motion with $E>0$ can occur in this case. In agreement with the aforementioned, for this motion, $r$ is bound from below by $r=r_{\rm min}$, satisfying the equality $U_{\rm eff}(r_{\rm min})=E$. It explains why the violation of Eq.~\eqref{eq:class_cond} makes collapse impossible.}

The case $\beta = \frac{M^2}{2m}$ is trivial since the dynamic equation, in this case, is the same as that for a free motion of the particle, when $m\dot{r}^2/2=E$. Therefore, in what follows, we suppose the strict inequality
\begin{equation}\label{eq:beta_class}
\beta > \frac{M^2}{2m}.
\end{equation}

First, we consider Eqs.~\eqref{eq:t(r)_class},~\eqref{eq:U} with {a finite negative $E$. Then, Eq.~\eqref{eq:t(r)_class} yields
\begin{equation}\label{eq:r(t)_class_finite_E_t<0}
  r=\sqrt{-\chi t \left(1+\frac{t}{T}\right)},\;\; -\frac{T}{2} \leq t \leq 0
\end{equation}
\begin{equation}\label{eq:r(t)_class_finite_E_t>0}
  r=\sqrt{\chi t \left(1-\frac{t}{T}\right)},\;\; 0\leq t \leq \frac{T}{2}.
\end{equation}
where
\begin{equation}\label{eq:chi}
  \chi \equiv \frac{2\sqrt{2m\beta - M^2}}{m},\;\;T=\frac{m\chi}{2|E|}.
\end{equation}
(we set $t_{\rm ini}=0$ and employ the initial condition \mbox{$r(0)=0$).} {Note that the dimension of $\chi$ is $length^2/time$.}}

{Branch Eq.~\eqref{eq:r(t)_class_finite_E_t<0} of the obtained solution describes a fall of the particle to the center, which begins with \mbox{$r=r_{\rm max}\equiv \sqrt{\chi T}/2$}, at \mbox{$t=-T/2$}, and ends at the origin of the coordinate system, at $t=0$. At $t>0$, the collapse terns into escape, described by branch Eq.~\eqref{eq:r(t)_class_finite_E_t>0}, so that $r(t)$ increases from 0, at $t=0$, to $r_{\rm max}$, at \mbox{$t=T/2$.}}

{Note also that $T$ and $r_{\rm max}$ are the only problem constants with the dimensions of time and length, respectively. When, during the collapse, $r(t)$ becomes much smaller than $r_{\rm max}$, the latter ceases to play the role of the problem characteristic scale. Moreover, the condition $r(t) \ll r_{\rm max}$ holds at $|t| \ll T$, see Eqs.~\eqref{eq:r(t)_class_finite_E_t<0},~\eqref{eq:r(t)_class_finite_E_t>0}. It means that, in this region, $T$ does not determine the characteristic temporal scale too. That is to say, close to the completion of the collapse (beginning of the escape), the problem loses its characteristic scales both in time and space. Then, according to the general principles of dimensional analysis~\cite{barenblatt1996,Sedov1993} the problem must become self-similar, {when its dynamic is described by a dimensionless ratio of $r$ to a certain power of $t$, instead of the two independent variables $r$ and $t$, in non-self-similar cases; see also below Sec.~\ref{sec:quant_coll}.} 

For the problem in question, it is so indeed: in the specified region, the term $t/T$ in Eqs.~\eqref{eq:r(t)_class_finite_E_t<0},~\eqref{eq:r(t)_class_finite_E_t>0} may be dropped. It transforms both branches into the self-similar solution}
{\begin{equation}\label{eq:r(t)_class_E=0}
  \frac{r}{\sqrt{\chi|t|}}=1,
\end{equation}
where the dimensionless $\xi = r/\sqrt{\chi|t|}$ may be regarded as a new self-similar quantity.}

{What happens if $E$ tends to zero from below? In this limit, both $T$ and $r_{\rm max}$ tend to infinity, and the periodic nonlinear oscillations of the particle become an aperiodic motion when, at $t<0$, the particle falls to the center from infinity and then, at $t>0$ escapes from the center, returning to infinity.    }

{Remarkably, that in this case, the self-similar solution Eq.~\eqref{eq:r(t)_class_E=0} becomes {\it exact}. At the same time, any general-type solution with finite $E<0$ is transformed into the self-similar one at the late stage of the collapse (the initial stage of the escape); {see Eqs.~\eqref{eq:r(t)_class_finite_E_t<0},~\eqref{eq:r(t)_class_finite_E_t>0}}. In other words, the exact self-similar solution Eq.~\eqref{eq:r(t)_class_E=0}, valid at $E=0$, is an {\it attractor\/} for any other solution with a finite $E$ exhibiting the collapse (escape). For this reason, the case $E=0$ will play a special role in the proceeding discussion; see Secs.~\ref{sec:quant_coll}--\ref{sec:examp}.}

{It is important to stress that close to the moment of the collapse completion or escape beginning, the transformation of a solution with any finite value of $E$ into a self-similar form is not a specific feature of the potential Eq.~\eqref{eq:U}. It is a generic property of any collapse-admitting potential with a power-type singularity. Indeed, consider a potential with the singularity $\sim 1/r^s$. To satisfy the collapse condition Eq.~\eqref{eq:class_cond} at  $M\neq 0$, we must have $s\geq 2$. However, if $M=0$, the only restriction is $s>0$. In this case, even the Coulomb potential $\sim 1/r$ is collapse-admitting. Either way, at $r\rightarrow 0$ and {\it any} finite $E$, the term $\sim 1/r^s$ makes the overwhelming contribution to the square root in Eq.~\eqref{eq:t(r)_class}. Then, dropping other terms under the square root sign and evaluating the integral, we obtain a universal self-similar solution in the form $r/|t|^{\frac{2}{2+s}} = const$, whose particular case at $s=2$ is Eq.~\eqref{eq:r(t)_class_E=0}. In other words, at $r \rightarrow 0$, the dependence  $r = const|t|^{\frac{2}{2+s}}$ is the only universal asymptotic to {\it any\/} collapse(escape)-exhibiting solution to the problem in question.}

Returning to the potential Eq.~\eqref{eq:U}, it is relevant to calculate the radial component of the particle momentum $p_r=mdr/dt$. At $E=0$, it reads:
\begin{equation}\label{eq:pr_class}
  p_r = \pm\frac{\chi m}{2\sqrt{\chi|t|}}%\;\;{\rm for}\;\;t<0\;\;{\rm and}\;\; p_r = \frac{\chi m}{2\sqrt{\chi t}}\;\;{\rm for}\;\;t>0,
\end{equation}
where the signs minus and plus correspond to the collapse ($t<0)$ and escape $(t>0)$, respectively. Note that $p_r$ diverges at $t\rightarrow 0$.

\begin{figure}[h]
  \centering
  \includegraphics[width=\columnwidth]{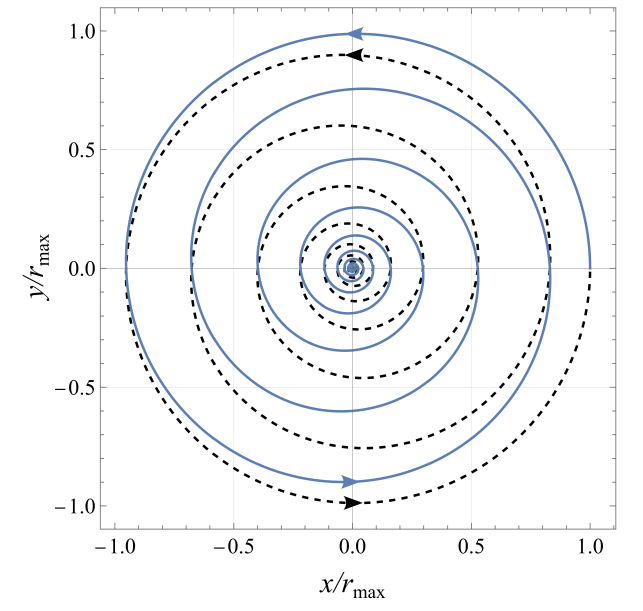}
  \caption{An example of a classical collapse-escape trajectory in a Cartesian coordinate system; $a=5$ . The dashed black and solid blue lines designate the two parts of the trajectory described by Eqs.~\eqref{eq:r(t)_class_finite_E_t>0},~\eqref{eq:phi(t)_finite_E} with $0\leq t \leq T$. Here \mbox{$0\leq t \leq T/2$} and $T/2 \leq t \leq T$ correspond to escape and collapse, respectively.}\label{fig:Spiral}
\end{figure}

{Taking into account the periodicity of the particle motion, we can rewrite the solution Eq.~\eqref{eq:r(t)_class_finite_E_t<0},~\eqref{eq:r(t)_class_finite_E_t>0} in an equivalent form, which sometimes is more convenient for analysis. To this end, we employ instead of Eq.~\eqref{eq:r(t)_class_finite_E_t<0}, the same branch shifted by period $T$ along the $t$-axis. It can be done with the help of the formal transformation $t \rightarrow t-T$ in Eq.~\eqref{eq:r(t)_class_finite_E_t<0}. It is easy to see that this procedure converts Eq.~\eqref{eq:r(t)_class_finite_E_t<0} into Eq.~\eqref{eq:r(t)_class_finite_E_t>0}. At the same time, the validity domain of the converted branch becomes $T/2\leq t \leq T$. Merging it with the validity domain stipulated by Eq.~\eqref{eq:r(t)_class_finite_E_t>0}, we obtain an equivalent form of a single period of the solution described only by Eq.~\eqref{eq:r(t)_class_finite_E_t>0}, which, however, now is valid for $0\leq t \leq T$.}

{To complete the inspection of the classical problem, we present an explicit expression for $\varphi(t)$. We readily obtain it by integrating Eq.~\eqref{eq:dphi/dt}. Doing that, it is convenient to use for $r(t)$ the above equivalent form of the solution. %, namely Eq.~\eqref{eq:r(t)_class_finite_E_t>0} with the $t$-domain extended to $0\leq t\leq T$.
It gives rise to the following formula:
\begin{equation}\label{eq:phi(t)_finite_E}
  \varphi(t) =  a\ln\frac{t}{T-t},\;\; 0 \leq t \leq T,
\end{equation}
where $a\equiv \frac{M}{m\chi}$. {We stress that, for the obtained solution, the domain of the non-trivial values of $\varphi$ extends from minus to plus infinity.} %The dependence of $\varphi(t)$ in the adjacent to $0 \leq t \leq T$ period $-T \leq t \leq 0$ follows from Eq.~\eqref{eq:phi(t)_finite_E} subjected to the transformation $t\rightarrow -t$; see Eq.~\eqref{eq:r(t)_class_finite_E_t<0}.}
%
%{In the self-similar limit (the vicinity of the moment $t=0$) these expressions are transformed into
%\begin{equation}\label{eq:phi(t)_E=0}
%\varphi(t) =  a\ln\frac{|t|}{T},
%\end{equation}
%Seemingly, Eq.~\eqref{eq:phi(t)_E=0} becomes meaningless at $E\rightarrow 0$, when the period of oscillations $T \rightarrow \infty$. However, it is not so. The point is that since the self-similar solution does not have any characteristic constant with the dimension of time, the role of $T$ in Eq.~\eqref{eq:phi(t)_E=0} may play any constant with the proper dimension. Its specific value does not affect the shape of the trajectory. Indeed, excluding $t$ from Eqs.~\eqref{eq:r(t)_class_E=0},~\eqref{eq:phi(t)_E=0} we obtain the explicit equation for the trajectory: $r=\sqrt{\chi T}\exp\left(\frac{\varphi}{2a}\right)$. If now we consider two arbitrary fixed values of \mbox{$T=T_{1,2}$,} then, at $T=T_2$, we have
%%
%\begin{equation*}
%  r\left|_{_{T_2}}\right.=\sqrt{\chi T_2}\exp\left(\frac{\varphi}{2a}\right)=\sqrt{\chi T_1}\exp\left(\frac{\varphi-\varphi_0}{2a}\right),
%\end{equation*}
%%
%where $\varphi_0=  a\ln(T_1/T_2)$. {Thus, we have obtained the same trajectory as that at $T=T_1$, only rotated by the angle $\varphi_0$.}   }
%
As an example, Fig.~\ref{fig:Spiral} shows the trajectory of the particle at $a=5$.} %The dashed and solid lines designate the parts of the trajectory corresponding to the collapse and escape, respectively.

\section{Fall of electron to heavy dipole: Radiative losses\label{sec:rad_loss} }

This section considers a specific example of the classical collapse: a fall of a (quasi)classical electron to a heavy neutral atom {(or molecule) with a fixed finite dipole moment $\mathbf{d}$~\footnote{{The Hamiltonian of an atom is invariant against the inversion transformation, while the Hamiltonian of a molecule is not. For this reason, a molecule may have a fixed dipole moment in a stationary state. In contrast, the dipole moment of an atom, in a generic case, is zero. Nonetheless, if certain special conditions hold, an atom also may have a finite dipole moment~\cite{landau2013quantum}.}}. We consider the most straightforward problem formulation, corresponding to the zero angular momentum of the electron. Then, $\mathbf{d}$ must be directed to the falling electron to minimize the electrostatic interaction energy. Thus, the problem is a particular case of the one discussed above, where $M=0$. We can neglect the displacement of the atom's center of mass owing to the tremendous difference in the masses of the electron and atom. However,}
%
%We can neglect the displacement of the atom's center of mass owing to the tremendous difference in the masses of the electron and atom. However, the dipole moment freely rotates to make the potential spherically symmetric. Then, $\mathbf{d}$ all the time must be directed to the falling electron to minimize the electrostatic interaction energy. Since the rotation of $\mathbf{d}$ occurs owing to the corresponding reorientation of the atom valence electrons and does not imply any rotation of the heavy nucleus, this assumption looks reasonable, at least as long as the angular velocity of the falling electron is not too large. {Alternatively, we can consider a fall of an electron with zero angular momentum. In this case, the reorientation of $\mathbf{d}$ is not required to fulfill the employed problem formulation. We are not presenting the corresponding criteria for these assumptions to hold. They are discussed in many quantum mechanics textbooks; see, e.g. Ref.~\cite{landau2013quantum}.} Our goal here is only to show how incorporating new effects into the problem formulation affects the obtained solutions.
%
the falling electron moves with an increasing acceleration and hence must emit electromagnetic waves. The emission decreases the electron energy. It may affect the structure of the above solutions. To elucidate this effect, we calculate the radiative losses. {We do it based on the above solution (obtained without consideration of the losses) regarding this solution as the zeroth approximation. Then,} the power emitted by a non-relativistic electron moving with the acceleration $w$ is given by Larmor's formula~\cite{landau2013classical}
\begin{equation}\label{eq:radiation_power}
  I=\frac{2e^2w^2}{3c^3},
\end{equation}
where $e$ is the electron charge, and $c$ stands for the speed of light in a vacuum.
%
%In the polar coordinate system, the acceleration vector $\mathbf{w}$ is described by two perpendicular components: radial ($w_r$) and transversal ($w_\varphi$) given by the following expressions:
%\begin{eqnarray}
%% \nonumber to remove numbering (before each equation)
%  w_r &=& \ddot{r} -r\dot{\phi}^2, \label{eq:wr} \\
%  w_\varphi &=& r\ddot{\phi}+2\dot{r}\dot{\varphi}, \label{w_phy}
%\end{eqnarray}
%
%However, employing the angular momentum conservation,  Eq.~\eqref{eq:dphi/dt}, it is easy to show that $w_\varphi \equiv 0$. Thus, at the motion in any central field $w = w_r$.

For the problem in question, the closer collapse completion moment, the larger $w$. Therefore, it is natural to suppose that the effect of the radiative losses increases, as the collapse completion approaches. On the other hand, as shown above, close to the completion moment, the general solution Eqs.~\eqref{eq:r(t)_class_finite_E_t<0},~\eqref{eq:r(t)_class_finite_E_t>0} is transformed into the self-similar form Eq.~\eqref{eq:r(t)_class_E=0}. Therefore, it is sufficient to study the radiative losses for \mbox{$r(t)=\sqrt{-\chi t}$} at \mbox{$t\leq 0$}.  In this case, the total energy $E_{\rm rad}$ emitted prior to a given moment $t$ is
\begin{equation}\label{eq:Erad}
  E_{\rm rad} = \int^{t}I(t')dt'=\frac{4e^2\beta^2}{3c^3m^2\chi(-\chi t)^2} \equiv \frac{4e^2\beta^2}{3c^3m^2\chi r^4}.
\end{equation}
The upper limit of the integral in Eq.~\eqref{eq:Erad} makes the main contribution to its value. Then, it is possible to extend the lower limit to minus infinity, despite the self-similar asymptotic Eq.~\eqref{eq:r(t)_class_E=0} of the general solution Eq.~\eqref{eq:r(t)_class_finite_E_t<0} is valid only in the vicinity of the collapse moment. %In addition, it verifies our assumption that the closer the moment of the collapse, the more pronounced the radiative losses.

To estimate how these losses affect the collapse dynamic, note that the latter is described by Eqs.~\eqref{eq:t(r)_class} and that close to the collapse completion moment \mbox{$|U(r)| \gg |E|$.} It means that we have to compare $E_{\rm rad}$ not with $E$ but with $U(r)$. In other words, the radiative losses impact on the collapse dynamic becomes substantial at the value of $r\sim r_{\rm \! rad}$, where $r_{\rm \! rad}$ is defined by the condition $|U(r_{\rm \! rad})|= E_{\rm rad}(r_{\rm \! rad})$. In contrast, the losses may be neglected at $r \gg r_{\rm \! rad}$. Simple algebra results in the following expression for $r_{\rm \! rad}$:
\begin{equation}\label{eq:r_rad}
  r_{\rm \! rad} = \frac{2|e|}{mc^{\frac{3}{2}}}\sqrt{\frac{\beta}{3\chi}}.
\end{equation}

Let us estimate the order of magnitude of $r_{\rm \! rad}$. At $M=0$, $\chi =2\sqrt{2\beta/m}$; see Eq.~\eqref{eq:chi}. Next, for the dipole moment produced by a polarization of spatial distribution of the valence electrons, $\beta$ is estimated as $e^2 r_{\!_B}$, where $r_{\!_B} = \frac{\hbar^2}{me^2}$ is Bohr's radius. In this case the estimate of the r.h.s. of Eq.~\eqref{eq:r_rad} reads
\begin{equation}\label{eq:r_rad_estimate}
  r_{\rm \! rad} \sim \frac{1}{\left(m r_{\!_B}\right)^{\frac{3}{4}}}\left(\frac{|e|}{c}\right)^{\frac{3}{2}}\!r_{\!_B}=\alpha^{\frac{3}{2}}r_{\!_B},
\end{equation}
where $\alpha = \frac{e^2}{\hbar c} \approx 1/137$ is the fine-structure constant. The numerical value of $\alpha^{\frac{3}{2}} \approx 6 \cdot 10^{-4}$.  That is to say, $r_{\rm \! rad} \ll r_{\!_B}$.

Since Bohr's radius is the characteristic quantum scale for atomic phenomena, the obtained estimate of $r_{\rm \! rad}$ means that {long} before the impact of the radiative losses on the collapse dynamics becomes noticeable, the classical description must be replaced by the corresponding quantum one. This essentially quantum description of the collapse is given below.

\section{Quantum problem formulation\label{sec:quant_coll}}

{ Conventionally, the Hamiltonian in  Schr\"{o}dinger's equation is a Hermitian operator with a complete set of orthogonal eigenfunctions. Accordingly, solutions to Schr\"{o}dinger's equation can be built as eigenfunction expansions. However, in the case of potential Eq.~\eqref{eq:U}, the eigenfunctions corresponding to different $E$ values are not necessarily orthogonal~\footnote{{The orthogonality may be imposed as an {\it additional\/} condition. It singles out a certain subset of eigenfunctions, while the eigenfunctions that do not satisfy the imposed condition are not orthogonal to the ones belonging to the subset. For details see Ref.~\cite{book:Morse_Feshbach}.}}. It is related to their behavior at $r\rightarrow 0$~\cite{book:Morse_Feshbach}. Since the orthogonality of Hamiltonian eigenfunctions with different values of $E$ is a direct consequence of its self-adjointness (see, e.g.~\cite{landau2013quantum}), the non-orthogonality, even for a single pair of them, means that the Hamiltonian with potential Eq.~\eqref{eq:U} {\it is not Hermitian}. It makes the possibility of building general solutions to the corresponding Schr\"{o}dinger's equation in the form of eigenfunction expansions questionable. At least, the author is unaware of any success in this way.}

{Regarding particular solutions, which may be built from the set of the Hamiltonian eigenfunctions discussed in the monograph by Morse and Feshbach~\cite{book:Morse_Feshbach}, they do not help much (if any) to understand the collapse dynamics. The point is that in  Schr\"{o}dinger's equation, Hamiltonian eigenfunctions are wave functions of {\it stationary states}. In the collapse case, building a solution in the form of eigenfunction expansion is an attempt to describe an essentially time-dependent process with the help of stationary-state eigenfunctions. It looks like trying to force a square peg into a round hole. To overcome this difficulty, we consider} a spatio-temporal evolution of a wave packet~\footnote{The idea is not mine. When I was a junior scientist, my adviser Sergei I. Anisimov from Landau Institute told me about this approach to the problem. Moreover, he said that together with Igor E. Dzyaloshinskii, they had found a solution describing the collapse of a wave packet.  Many years later, in connection with nanoparticle light scattering, I came across a problem mathematically analogous to the quantum
collapse. I recalled this conversation with Anisimov and asked him for details and references. He replied that these results had never been published and details he did not remember. Now, Anisimov and Dzyalashinskii have both passed away. I decided to apply their approach to the problem and make the results available to a broad readership as a small token of my great respect for these distinguished scholars.}, described by time-dependent Schr\"{o}dinger's equation:
\begin{equation}\label{eq:Schr}
 i\hbar \frac{\partial \Psi}{\partial t} = \hat{H}\Psi;\;\; \hat{H} \equiv -\frac{\hbar^2}{2m}\Delta + U(r).
 %\vspace*{6pt}
\end{equation}
%\mbox{}\\
Here $\Delta$ stands for the Laplacian, and $U(r)$ is given by Eq.~\eqref{eq:U}. Equation~\eqref{eq:Schr} should be supplemented by the initial condition $\Psi = \Psi_0(\mathbf{r})$ and the one stipulating finiteness of the wave function norm, i.e, convergence of $\int |\Psi|^2d^3\mathbf{r}$.

We employ the power of dimensional analysis to obtain collapsing solutions of Eq.~\eqref{eq:Schr} The main concepts of this analysis are as follows~\cite{Sedov1993,barenblatt1996}. Arguments of any mathematical function must be dimensionless quantities. The wave function in Eq.~\eqref{eq:Schr} depends on the dimensional $\mathbf{r}$ and $t$. To make them dimensionless, we have to normalize them by constants with the proper dimensions. These constants should be built from the ones entering our problem. There are only three dimensional constants entering Eq.~\eqref{eq:Schr}, namely $m,\;\hbar,$ and $\beta$. Their most general combination has the form: $m^{\zeta_1}\hbar^{\zeta_2}\beta^{\zeta_3}$, where $\zeta_{1,2,3}$ are real constants. It is easy to see that for the $\beta$ dimension corresponding to Eq.~\eqref{eq:U} and any values of $\zeta_{1,2,3}$ this product cannot have the dimensions of $\mathbf{r}$ or $t$.

{The only remaining way to obtain the required constants is to find them in the initial condition. Indeed, the initial condition must have a characteristic spatial scale of the wave function localization {$r_{\rm ini}$. Then, e.g. the combination $m r_{\rm ini}^2/\hbar$ may be selected as the characteristic temporal scale.} Seemingly, it removes the normalization problem. It does not! The point is that if the collapse indeed takes place, the region of the wave function spatial localization contracts. Its size eventually turns to zero. Then, the finite scale associated with the initial conditions ceases to play the role of a characteristic scale of the problem and becomes useless for our purposes.}

{The other striking conclusion following from the dimensional analysis is that Schr\"odinger's equation with potential in the form of Eq.~\eqref{eq:U} cannot have discrete levels but the one with $E=0$~\cite{guggenheim1966inverse}. To see that, once again we consider a more general form of the potential, namely 
\begin{equation}\label{eq:U_s}
  U(r) = -\frac{\beta}{r^s},
\end{equation}
where the sign of $s$ may be any. At $s<0$ to make expression \eqref{eq:U_s} a potential well (not a barrier) $\beta$ also should be negative. In particular, at $s=1$ Eq.~\eqref{eq:U_s} corresponds to the Coulomb field, and for $s=-2$ it is a harmonic oscillator potential. In our case $s=2$. According to precisely the same arguments as those used above in the discussion of the characteristic spatio-temporal scales, we conclude that if a discrete spectrum exists, its levels should be composed as products of powers of $\hbar,\;m$ and $\beta$. There is the only possible product with the required dimension, which gives rise to the following expression for the energy of the levels:
\begin{equation}\label{eq:E_n}
  E_n=\epsilon_n m \left(\frac{|\beta| m^{s-1}}{\hbar^s}\right)^{\frac{2}{2-s}},
\end{equation}
where $\epsilon_n$ is a dimensionless quantity.}

{It is seen straightforwardly that Eq.~\eqref{eq:E_n} coincides with the well-known expressions for the spectrum of a harmonic oscillator and the one in the Coulomb potential. However, at $s=2$ it fails to produce a reasonable result. In this case, the only opportunity to return to physically meaningful application of the obtained expression is to set there $\epsilon_n$ to zero.}

The above simple arguments result in the conclusion that if the quantum collapse exists, a solution describing its final stage must be self-similar. In this solution, $\mathbf{r}$ varies as a certain power of $t$, {and the corresponding dimensionless variable is built as a ratio of $\mathbf{r}$ to this power of $t$, cf. Eq.~\eqref{eq:r(t)_class_E=0} in the classical case}. Moreover, the self-similar solution must be an attractor: any solution exhibiting the collapse is transformed into the self-similar form at the final stage of the phenomenon. {We have proven that for the classical case; see the convergence of the general solution Eqs.~\eqref{eq:r(t)_class_finite_E_t<0},~\eqref{eq:r(t)_class_finite_E_t>0} to the self-similar Eq.~\eqref{eq:r(t)_class_E=0} in the vicinity of the point $r=0$. We can transfer these classical results to the quantum problem, matching the classical and quantum cases through the quasi-classical approximation, discussed below.}

{\section{Quasi-classical condition}\label{sec:quasi-class}} 

{The applicability condition for the quasi-classical approximation implies that the characteristic spatial scale of the wave function variations is much smaller than that for the potential and hence, for the corresponding classical solution. The particle de Broglie wavelength, which is inversely proportional to its momentum, determines the wave function spatial variations. At a given potential, the larger the momentum, the smaller the de Broglie wavelength and, therefore, the more accurate the quasi-classical approximation. For the classical problem formulation, the particle momentum increases with the increase in the coupling constant $\beta$. On the other hand, the shape of the potential Eq.~\eqref{eq:U} is determined by $r^2$ in the denominator and does not depend on $\beta$. Then, it may be expected that the entire dynamic of the quantum particle is quasi-classical, at large enough $\beta$.}

{To check the guess, we must explicitly employ the applicability condition. Owing to the introduced $U_{\rm eff}$ the classical particle dynamic becomes one-dimensional, see Eq.~\eqref{eq:E}--\eqref{eq:t(r)_class}, and the radial component of the momentum $p_r = m\dot{r}$ plays the role of the corresponding one-dimensional momentum. In this case, the quasi-classical applicability condition reads as follows~\cite{landau2013quantum}:
\begin{equation}\label{eq:q-cl.applic.landau}
  \left|\frac{\partial \lambdabar }{\partial r}\right| \ll 1,
\end{equation}
where $\lambdabar = \hbar/p_r$ and {$p_r=\pm \sqrt{2m\left(E-U_{\rm eff}\right)}$; see Eq.~\eqref{eq:E}.}
}

{Simple calculations transform Eq.~\eqref{eq:q-cl.applic.landau} into the following expression:
\begin{equation}\label{eq:q-cl.E<0}
  \frac{\hbar(2m\beta -M^2)}{|2mEr^2+2m\beta-M^2|^{3/2}} \ll 1.
\end{equation}
Thus, at any finite $E$ and $r \rightarrow \infty$, the particle motion is quasi-classical. However, we are interested in the opposite limit, namely $r \rightarrow 0$. Setting in Eq.~\eqref{eq:q-cl.E<0} $r$ to zero gives rise to the condition
\begin{equation}\label{eq:q-cl.E=0}
  \frac{\hbar}{\sqrt{2m\beta - M^2}} \ll 1.
\end{equation}
Eq.~\eqref{eq:q-cl.E=0} mathematically confirms the guess: at large enough $\beta$, the quantum dynamic is always quasi-classical.  }
\mbox{}
{\section{Exact solution to Schr\"{o}dinger's equation}\label{sec:wave_function}}

{Now, when we have unveiled the qualitative features of the quantum problem, we can find its exact self-similar solution. We will look for it in the following form:}
{\begin{equation}\label{eq:Self_Psi}
  \Psi = \sum_\ell\sum_{m=-\ell}^{\ell}C_{\ell m}\Phi_\ell(-\chi t) Y_\ell^m(\theta,\varphi)R_\ell(\xi).
\end{equation}
%\mbox{}\\
Here $C_{\ell m}$ are constants, $Y_\ell^m(\theta,\varphi)$ stand for the spherical harmonic functions (do not confuse index $m$ with the mass of the particle), $\xi = r/(-\chi t)^\nu$; $\Phi_\ell(-\chi t),\;R_\ell(\xi)$, $\chi, \;\nu$ are yet unknown functions and constants, respectively. Here the moment $t=0$ corresponds to the complete collapse. Then, the collapse dynamic is described by $t<0$.

{We have to stress that, in contrast to the conventional eigenfunction expansion, the ansatz Eq.~\eqref{eq:Self_Psi} {\it is not\/} a general solution to the problem. Moreover, for the time being, we even cannot say that it is a solution. We {\it guess} that it is. To ensure that the solution in such a form does exist, we have to find it explicitly. Let us proceed in this way.}

It is possible to show (see Appendix \ref{app:self-sim}) that Eq.~\eqref{eq:Self_Psi} may be a self-similar solution to Eq.~\eqref{eq:Schr}, provided
\begin{equation}\label{eq:Phi_ell}
  \Phi_\ell(-\chi t) = const_\ell(-\chi t)^{\mu_\ell},
\end{equation}
where $\mu_\ell$ are dimensionless constants (generally speaking, complex). Regarding the dimensional constant $\chi$, it is convenient to suppose that \mbox{$\chi = \hbar/m$.} {Note that, at this definition, the dimension of $\chi$  is the same as that in the classical case, namely $length^2/time$.}

{The necessary collapse {condition} \mbox{$\ell(\ell+1)< \frac{2m\beta}{\hbar^2}-\frac{1}{4}$~\cite{landau2013quantum,book:Morse_Feshbach}}, limits the value of $\ell$ in the sums in Eq.~\eqref{eq:Self_Psi}. Recalling that the eigenvalues of the square of the angular momentum $\widehat{\mathbf{l^2}}$ are $\hbar^2\ell(\ell+1)$ and denoting them as $M^2$, we may rewrite the above constraint as
{
\begin{equation}\label{eq:beta_quat}
  \beta >\frac{\hbar^2\ell(\ell+1)}{2m}+\frac{\hbar^2}{8m} \equiv \frac{M^2}{2m}+\frac{\hbar^2}{8m}.
\end{equation}
When $\hbar \rightarrow 0$, Eq.~\eqref{eq:beta_quat} coincides with the classical collapse condition; see Eq.~\eqref{eq:beta_class}.}

Substituting Eqs.~\eqref{eq:Self_Psi},~\eqref{eq:Phi_ell} into Eq.~\eqref{eq:Schr} and employing the orthogonality of $Y_\ell^m(\theta,\varphi)$ we obtain a {detached} equation for $R_\ell$. Then, without loss of generality, we {consider} a single term on the r.h.s. of Eq.~\eqref{eq:Self_Psi} as $\Psi$, dropping the signs of sums, while $C_{\ell m}$ may be dropped owing to the linearity of the problem. Therefore, we can simplify the notations by omitting the subscript $\ell$. Eventually, at $\nu = 1/2$, Schr\"{o}dinger's equation is reduced to the following ordinary differential equation for $R(\xi)$; see Appendix~\ref{app:self-sim}:
\begin{equation}\label{eq:R}
  R''+\left(\frac{2}{\xi}+i\xi\right)R'+\left(\frac{\gamma}{\xi^2}-2i\mu\right)R=0,
\end{equation}
where
\begin{equation}\label{eq:gamma}
\gamma \equiv \frac{2m\beta}{\hbar^2}-\ell(\ell+1),
\end{equation}
and prime denotes $d/d\xi$.

Zakharov \& Kuznetsov~\cite{zakharov1986_collapse_JETP,zakharov2012_Collapse_UFN} employed a particular type of solution Eqs.~\eqref{eq:Self_Psi},~\eqref{eq:Phi_ell} with $\mu = -(1/2+i\kappa)$ to study the so-called weak collapse in {\it nonlinear} Schr\"{o}dinger's equation. However, in Refs.~\cite{zakharov1986_collapse_JETP,zakharov2012_Collapse_UFN} a physically meaningful solution exists only at a single value of $\kappa$. By contrast, in the linear problem discussed here, the restrictions imposed on $\mu$ are much weaker, see below.

{Equation~\eqref{eq:R}} is exactly integrable. Its general solution is
\begin{eqnarray}
  & &\!\!\!\!\!\!  R(\xi) =\nonumber\\
  & & \!\!\!\frac{1}{\sqrt{\xi}}\left[C_1 \xi^{-\frac{i \alpha }{2}}\!\!\,_1\!F_1\left(-\frac{1+i \alpha }{4}-\mu ;1-\frac{i \alpha }{2};-\frac{i\xi ^2}{2} \right)\right.  \nonumber\\
  & & \!\!\! \left.+\, C_2 \xi^{\frac{i \alpha }{2}}\!\!\,_1\!F_1\left(-\frac{1-i \alpha }{4}-\mu ;1+\frac{i \alpha }{2};-\frac{i\xi ^2}{2} \right)\right].\label{eq:Sol_R}
\end{eqnarray}
Here $C_{1,2}$ are constants, $_1F_1(a;b;z)$ designates the Kummer confluent hypergeometric function of the first kind~\cite{janke1960tafeln}, and
\begin{equation}\label{eq:alpha}
  \alpha \equiv \sqrt{4\gamma - 1}>0,
\end{equation}
(do not confuse it with the fine-structure constant!). 

The positiveness of the expression under the square root in Eq.~\eqref{eq:alpha} {follows from the necessary collapse condition; see Eqs.~\eqref{eq:beta_quat},~\eqref{eq:gamma}. If the condition does not hold, i.e. \mbox{$\gamma <1/4$,} $\alpha$ is purely imaginary, and %the expression for $R(\xi)$ in the form of 
Eq.~\eqref{eq:R} with purely real $\alpha$ becomes invalid.}

{Now, when we have introduced the self-similar variable  $\xi$, and obtained the quantum solution Eq.~\eqref{eq:Sol_R},  we can rewrite Eq.~\eqref{eq:q-cl.E=0} in an equivalent form, namely
\begin{equation}\label{eq:q-cl.r_q/r_c}
  2\chi|t|/r_{\rm clas}(t)^2 \ll 1.
\end{equation}
Here $r_{\rm clas}(t)$ is the classical self-similar solution Eq.~\eqref{eq:r(t)_class_E=0}. The advantage of this presentation is its physical clarity. Indeed, $\sqrt{-\chi t}$ is the characteristic spatial scale of variations of the obtained self-similar wave function. Then, the quantum particle motion is quasi-classical, provided this scale is small relative to the one for the corresponding classical solution. Notably, while the condition Eq.~\eqref{eq:q-cl.r_q/r_c} is valid for any general (i.e. not {\it necessarily} self-similar) solution, it is written in terms of the self-similar variables. The latter is an additional indication of the importance of self-similar solutions for the problem in question.}

Expression~\eqref{eq:Sol_R} includes the terms $\xi^{\pm i\frac{\alpha}{2}}$, where $\xi$ and $\alpha$ are positive quantities. Then,
\begin{eqnarray}
  \xi^{\pm i\frac{\alpha}{2}}&=&(e^{i2\pi n+\ln \xi})^{\pm i\frac{\alpha}{2}} = e^{\mp\pi\alpha n}e^{\pm i\frac{\alpha}{2}\ln \xi} \label{eq:xi^ia} \\
  &=&e^{\mp\pi\alpha n}\left[\cos\left( \frac{\alpha}{2}\ln \xi\right) \pm i\sin\left(\frac{\alpha}{2}\ln \xi\right)\right],\nonumber
\end{eqnarray}
where $n$ is {an} arbitrary integer. {Such a singularity is typical for the problem under consideration~\cite{landau2013quantum,book:Morse_Feshbach}.}

Expression~\eqref{eq:xi^ia} has an infinite number of branches corresponding to different values of $n$. Every branch's real and imaginary parts have the number of zeros, {demonstrating unlimited growth at} $\xi \rightarrow 0$. For simplicity, in what follows, only the single branch with $n=0$ is inspected.

{According to the definition of $\xi$, we should emphasize that this quantity diverges at $t=0$, i.e. at the moment of the collapse completion. Then, while the behavior of $R(\xi)$ at $\xi \rightarrow 0$ is essential from the viewpoint of the solution branching and its other analytical properties, the behavior of the wave function close to the moment of the collapse completion is practically overwhelmingly determined by the $R(\xi)$ asymptotic at $\xi \rightarrow \infty$. This asymptotic is discussed in Appendix~\ref{app:norm}.     }

{What is about escape? Schr\"{o}dinger's equation is invariant against the time-reversal procedure accompanied by the complex conjugation. Then, being applied to \mbox{Eqs.~\eqref{eq:Self_Psi},~\eqref{eq:Phi_ell},~\eqref{eq:Sol_R}}, these transformations generate the wave function describing the particle escape from the center at $t>0$. As well as in the classical case, at the moment $t=0$, the collapse is transformed into escape by transferring from one solution to the other.}

%{In the end of this section note, that for any finite $r$ the solution Eq,~\eqref{eq:R} does not exist at $t=0$, i.e. at the very moment of the collapse. It resolves the mentioned in the {\it Introduction} contradiction between the possibility for the wave function localization area to collapse to a zero-dimensional singular point and the nonexistence of the wave function limit at $r$ tending to the same singularity.}
\mbox{} 
{\section{Time-dependent Norm}\label{sec:norm}}

We must normalize $\Psi$ to calculate the probability density and the mean values of operators. Conventionally, the norm of a wave function ($\|\Psi\|$) is a constant. It follows from self-adjointness of $\hat{H}$~\cite{landau2013quantum}. {Indeed,
\begin{eqnarray}
   & & \frac{d}{dt}\|\Psi\|^2 \equiv \frac{d}{dt}\langle\Psi|\Psi\rangle \equiv \frac{d}{dt}\int\Psi^\ast\Psi d^3\mathbf{r} \label{eq:dN/dt_1}\\
   & & = \int \Psi \frac{\partial\Psi^*}{\partial t}d^3\mathbf{r}+\int \Psi^* \frac{\partial\Psi}{\partial t}d^3\mathbf{r},\nonumber
\end{eqnarray}
where the asterisk stands for the complex conjugation. Bearing in mind Eq.~\eqref{eq:Schr}, %and the fact that $\hat{H}^* = \hat{H}$,
it may be rewritten as
\begin{equation}\label{eq:dN/dt_2}
  \frac{i}{\hbar}\left(\int \Psi\hat{H}^*\Psi^* d^3\mathbf{r}-\int \Psi^*\hat{H}\Psi d^3\mathbf{r} \right).
\end{equation}
If $\hat{H}$ is Hermitian, the above expression identically equals zero~\footnote{{Actually, the case is more subtle. In Eqs.~\protect \eqref {eq:dN/dt_1},~\protect \eqref {eq:dN/dt_2} we integrate over all space at once. A more accurate approach implies that we take the integrals over some finite volume $V$, and then, consider the limit $V\rightarrow \infty $, extending $V$ to all space. For the finite $V$, the integrals over $V$ can be transformed by Gauss's theorem into integrals over the bounding $V$ surfaces. The corresponding integrals are proportional to the probability density fluxes through these surfaces. The expression Eq.~\protect \eqref{eq:dN/dt_2} vanishes, provided these fluxes turn to zero at $V\rightarrow \infty $  or, in a more general case, the flux from infinity is equal to the one through the origin of the coordinate system. See the discussion of this issue in Sec.~\ref{sec:contin}.}}.}

In our case, $\hat{H}$ is not Hermitian. Then, the norm may be time-dependent. Indeed, simple calculations show that for Eqs.~\eqref{eq:Self_Psi},~\eqref{eq:Phi_ell},~\eqref{eq:Sol_R} \mbox{$\|\Psi\|^2 = C(-\chi t)^{\frac{3}{2}+2\mu'}$}, where $\mu'={\rm Re}\,\mu$ and $C$ is a constant proportional to $\int_{0}^{\infty}\left|R(\xi)\right|^2 \xi^2d\xi$. The convergence of the latter determines the finiteness of~$C$.

Calculations show that a proper choice of the ratio $C_{2}/C_1$ in Eq.~\eqref{eq:Sol_R} provides the integral convergence at any values of $\mu'$ except \mbox{$\mu'=-3/4$}.  At \mbox{$\mu'=-3/4$} the integral diverges at any $C_{1,2}$; see Appendix~\ref{app:norm}. Since \mbox{$\mu'=-3/4$} is the only value of $\mu'$, when $\|\Psi\|$ would not depend on $t$ (see above), the requirement of the finiteness of $\|\Psi\|$ {\it always} makes it time-dependent.

{Note that the divergence of $\|\Psi\|^2$ at \mbox{$\mu'=-3/4$} is logarithmic, i.e. extremely weak. It can be stabilized by any correction making the wave function decay faster than that for the obtained exact solution. The role of this correction can play, e.g. relativistic effects, the radiative losses discussed above, etc. Phenomenologically the stabilization may be introduced as a cut-off of the norm integral at some $\xi_0 \gg 1$. In this case, the norm is a constant, and all further consideration is the same as that in conventional cases with a Hermitian Hamiltonian. Then, we could assume that the case \mbox{$\mu'=-3/4$} is the only physical one, while all other solutions with \mbox{$\mu'\neq-3/4$} do not have physical meaning.}

{However, the cut-off drives the problem beyond its initial, strict all-sufficient formulation of Schr\"{o}dinger's equation with potential Eq.~\eqref{eq:U} defined in all unlimited space. It is interesting to see if we can keep the problem formulation and the basic quantum mechanics concepts unmodified for a wave function with a time-dependent norm. {We discuss this issue in the next section.}}
\mbox{}\\
{\section{probability density and mean value calculations% for wave function with time-dependent norm
}\label{sec:mean_values}}
{To be able to employ a wave function with a non-conserved norm we must reexamine several conventional quantum mechanics rules and modify them, if required. The first, arising in this case question,} is how to normalize $\Psi$ to obtain the probability density? There are at least two options:
%\vspace*{-6pt}
    \begin{itemize}
      \item[(i)] the conventional expression, namely $|\Psi|^2/const$; \vspace*{-6pt}
      \item[(ii)] $|\Psi|^2/\|\Psi\|^2$.
    \end{itemize}
%\vspace*{-4pt}
{Both have {\it pro} and {\it contra}. Case (i) is conventional. However, in this case, the probability of finding a particle in any point of all space is not equal to unity and varies in time as  $(-\chi t)^{\frac{3}{2}+2\mu'}$ for collapse and as $(\chi t)^{\frac{3}{2}+2\mu'}$ for escape, see the previous section. Then, values of $\mu'$ smaller than $-3/4$ are meaningless since they would correspond to the probability of finding the particle larger than unity at $|t|\rightarrow 0$. At $\mu'>-3/4$, (i) would mean that the singularity at $r=0$ acts as a sink for the collapse ($t<0$) and a source for the escape ($t>0$). In other words, during the collapse, the particle gradually gets out of our world (where to?), completely disappears at $t=0$, then, slowly returns at $t>0$. This scenario sounds rather unusual.}

{Though case (ii) is also unusual owing to the time-dependence of $\|\Psi\|$, in this case, the probability of finding a particle in any point of all space identically equals unity, and the problem of communication with  the ``other world" does not arise. However, in this case, the normalized wave function does not satisfy Schr\"{o}dinger's equation (the term $i\hbar\Psi\partial(1/\|\Psi\|)/\partial t$ remains uncompensated.) On the other hand, since $\Psi$ does, this feature does not contradict to the fundamentals of quantum mechanics. Thus, case (ii) looks more physical.}

{Nonetheless, the final judgement in favor of either of the two cases must be done with the help of the {\it uncertainty relations}. The derivation of the latter does not employ the self-adjointness of Hamiltonian and norm conservation (see Landau \& Lifshitz~\cite{landau2013quantum} and Appendix~\ref{app:uncert}). Therefore, the uncertainty relations must} be valid {for} the problem in question too. Calculating the mean value of $\hat{r}$ and $\hat{p}_r \equiv -i\hbar\partial/\partial r$, and assuming, as usual, that $\Delta r\Delta p_r \sim \langle \hat{r}\rangle\langle\hat{p}_r\rangle$, we readily obtain: in case (i) $\Delta r\Delta p_r \sim \hbar (-\chi t)^{3+4\mu'}$, which does not satisfy the uncertainty relations; in case (ii) $\Delta r\Delta p_r \sim \hbar$, which does. Thus, choice (ii) is correct, while (i) is not.

The angular momentum conservation gives an additional argument supporting the choice of case (ii). Owing to the problem symmetry, the mean angular momentum $\langle \widehat{\mathbf{l^2}}\rangle$ must be a conserved quantity. For the given $\Psi$, we have $\langle \Psi|\widehat{\mathbf{l^2}}|\Psi\rangle = \hbar^2\ell(\ell+1)\|\Psi\|^2$. {To avoid misunderstanding, we stress that here and in what follows, for any operator $\hat{A}$, the expression  $\langle \hat{A}\rangle$  designates the mean value of $\hat{A}$, while $\langle \Psi|\hat{A}|\Psi\rangle$ stands for the scalar product of $\langle \Psi|$ and $\hat{A}|\Psi\rangle$, where, generally speaking, the wave functions are not normalized. Therefore, $\langle \hat{A}\rangle = \langle\Psi|\hat{A}|\Psi\rangle/\|\Psi\|^2$  }

{Bearing this in mind and taking into account that $\|\Psi\|$ is time-dependent, we conclude that to make $\langle \widehat{\mathbf{l^2}}\rangle$  conserved, one must get rid of $\|\Psi\|$ in the expression for $\langle \widehat{\mathbf{l^2}}\rangle$, i.e. choose case (ii). Then, \mbox{$\langle \widehat{\mathbf{l^2}}\rangle = \hbar^2\ell(\ell+1) = const$}, as it should be.}       

It is also relevant to mention that in case (ii)
\begin{equation}\label{eq:<r>&<p>}
 \langle\hat{r}\rangle =
  c_{r}\sqrt{-\chi t};\;\langle\hat{p}_r\rangle =
   c_{p}\frac{\hbar}{\sqrt{-\chi t}},
\end{equation}
where $c_{r,p}$ are dimensionless constants of the order of unity. These expressions are remarkably similar to their classical analogs; see Eqs.~\eqref{eq:r(t)_class_E=0},~\eqref{eq:pr_class}.

Regarding the particle energy, note that since the obtained $\Psi$ is not an eigenfunction of $\hat{H}$, only the mean value: $E=\langle \hat{H} \rangle$ makes sense. According to Eq.~\eqref{eq:Schr} \mbox{and (ii),}
\begin{equation}\label{eq:<E>}
 E= \langle \Psi|\hat{H}|\Psi\rangle/\|\Psi\|^2 = i\hbar\langle \Psi|\partial\Psi/\partial t\rangle/\|\Psi\|^2 = C_E/t.
% \vspace*{12pt}
\end{equation}
Here the constant $C_E$ is given by a certain integral. The integral converges at $\mu'<-3/4$ and $\mu'>-1/4$; see Appendix \ref{app:norm}. {On the other hand, the energy conservation law requires that $E=const$. It is compatible with Eq.~\eqref{eq:<E>} only if  $C_E=0$, i.e. $E=0$.} {Tedious evaluation of the integral in the complex plane gives rise to the same result}; {see also the discussion at the end of Sec.~\ref{sec:quant_coll} concerning the non-existence of the discrete spectrum but the level with $E=0$ for the potential Eq.~\eqref{eq:U}, which also can be applied to this case.}

We stress that the condition $E=0$ also follows from the self-similarity of the obtained solution. {This argument is valid both in classical and quantum cases. Indeed, at $E\neq 0$, the problem possesses constants with the dimensions of time and length, which can be built with the help of  $m,\;\beta$ and $|E|$; see Eqs.~\eqref{eq:chi},~\eqref{eq:r(t)_class_E=0}. In the quantum case, due to the additional constant $\hbar$, it can be done even in various ways. The existence of a characteristic spatio-temporal scale breaks self-similarity, and expressions Eqs.~\eqref{eq:r(t)_class_E=0},~\eqref{eq:Self_Psi} fail to be {\it exact} solutions to the corresponding problems. It does not mean that, at $E\neq 0$, collapsing solutions do not exist at all. They may exist in non-self-similar forms; see Eqs.~\eqref{eq:r(t)_class_finite_E_t<0},~\eqref{eq:r(t)_class_finite_E_t>0}.} However, as we have several times emphasized above, the self-similar solutions remain attractors to non-self-similar ones.} %As we have shown discussing the classical collapse, they exist but in a non-self-similar form; see Eqs.~\eqref{eq:r(t)_class_finite_E_t<0},~\eqref{eq:r(t)_class_finite_E_t>0}. Nonetheless, the self-similar solution Eqs.~\eqref{eq:r(t)_class_E=0} remains an attractor for the non-self-similar ones. Since the dimensional analysis arguments are insensitive to the   }
\vspace*{-12pt}

{\section{Generalized Continuity equation\label{sec:contin}}}

How do the peculiarities mentioned above affect the continuity equation? To answer the question, we revise the corresponding conventional case~\cite{landau2013quantum}. According to it, the wave function satisfies the continuity equation
\begin{equation}\label{eq:continuity}
  \partial \left|\Psi\right|^2/ \partial t + {\rm div}\, \mathbf{j} = 0,
\end{equation}
where $\mathbf{j} \equiv \frac{i\hbar}{2m}\left(\Psi\nabla\Psi^*-\Psi^*\nabla\Psi\right)$ is the probability current density.

Equation~\eqref{eq:continuity} is a direct consequence of Schr\"{o}dinger's equation and the identities
\begin{equation}\label{eq:identities}
  \hat{H} \equiv \hat{H}^*;\;\; \Psi\Delta\Psi^*-\Psi^*\Delta\Psi \equiv {\rm div}\,(\Psi\nabla\Psi^*-\Psi^*\nabla\Psi)
\end{equation}

Conditions~Eq.~\eqref{eq:identities} hold for the problem in question, and hence, Eq.~\eqref{eq:continuity} is valid in this case too. However, now, due to the time dependence of the norm and (ii), neither $|\Psi|^2$ is the probability density, nor $\frac{i\hbar}{2m}\left(\Psi\nabla\Psi^*-\Psi^*\nabla\Psi\right)$ is the probability density current. The corresponding quantities are $\rho=|\Psi|^2/\|\Psi\|^2$ and \mbox{$\mathbf{J}=\frac{i\hbar}{2m \|\Psi\|^2}\left(\Psi\nabla\Psi^*-\Psi^*\nabla\Psi\right)$}, respectively.
Then, instead of Eq.~\eqref{eq:continuity} the continuity equation reads
\begin{equation}\label{eq:continuity_N}
  \frac{\partial\rho}{\partial t} + {\rm div}\, \mathbf{J} = -\frac{\rho}{\|\Psi\|^2}\frac{d\|\Psi\|^2}{dt} = -\left(\frac{3}{2}+ 2\mu'\right)\frac{\rho}{t}.
\end{equation}
To write the last equality in Eq.~\eqref{eq:continuity_N} we have employed the expression  $\|\Psi\|^2 = C(-\chi t)^{\frac{3}{2}+2\mu'}$.

If we integrate Eq.~\eqref{eq:continuity_N} over a certain closed volume $V$, the integral of $\partial\rho/\partial t$ gives the rate of the probability variation of finding the particle in this volume. {The integral of $ {\rm div}\, \mathbf{J}$ is transformed into the integral over the bounding $V$ surface(s): it is the flux of the probability density current through this surface(s).}

In the conventional case, the r.h.s. of the continuity equation is zero~\cite{landau2013quantum}, and the flux is the only cause of the probability variations. In contrast, since Eq.~\eqref{eq:continuity_N} has a non-zero r.h.s., the latter also contributes to the probability variations. Notably, the meaning of this contribution is the generation or leakage (depending on the sign of the r.h.s.) of the probability density inside~$V$.

Seemingly, this feature corresponds to creating something from nothing (generation) or vice versa (leakage). However, it does not! Such a behavior directly follows from the variations of the {\it size} of the wave function spatial localization region due to the collapse (escape). Indeed, let us consider a particle in a square potential well with infinitely high walls, as an example. Suppose, these walls ``adiabatically" move either toward or opposite each other. Then, obviously, the probability of finding the particle in a given part of the space inside the well changes in time, while the probability flux through the walls is zero: the changes are caused solely by the variations of the size of the wave function localization region. The same effect takes place during the collapse (escape).

Now, we integrate Eq.~\eqref{eq:continuity_N} over all space. Since, by definition, in this case, $\int \rho d^3\mathbf{r} \rightarrow 1$, Eq.~\eqref{eq:continuity_N} yields
\begin{equation}\label{eq:continuity_N_infty}
  \int {\rm div}\,\mathbf{J} d^3\mathbf{r} = -\left(\frac{3}{2}+ 2\mu'\right)\frac{1}{t}.
\end{equation}
Regarding the integral on the l.h.s. of Eq.~\eqref{eq:continuity_N_infty}, placing two concentric spheres about the point $r=0$ and employing Gauss's theorem, we reduce the integral over the volume to the ones over the surfaces of the spheres. Then, we let the inner and outer spheres' radii tend to zero and infinity, respectively.
Calculations based on the asymptotical expressions for $R(\xi)$ at \mbox{$\xi \rightarrow 0$} and \mbox{$\xi \rightarrow \infty$} presented in Appendix~\ref{app:norm} show that, in this case, the flux through the outer sphere tends to zero, while for the inner sphere, this is not the case, namely the singularity at $r=0$ acts as a sink, at \mbox{$\mu'>-3/4$}, and as a source, at \mbox{$\mu'<-3/4$}, in entire agreement with the expression   \mbox{$\|\Psi\|^2 = C(-\chi t)^{\frac{3}{2}+2\mu'}$.}

{It is important to stress that existence of the sink (source) at $r=0$ does not mean the particle gradually gets out (in) our world to (from) the singularity. Owing to the selected normalization rule (ii), the probability of finding the particle in all space remains fixed and always equals unity.}
%\vspace*{12pt}

\section{Specific Examples\label{sec:examp}}

To {illustrate} the behavior of the obtained solutions, we consider specific examples of Eq.~\eqref{eq:Sol_R} at $\mu=0$,
\begin{eqnarray}
  C_1 &=& \left(\frac{i}{2}\right)^{-\frac{i \alpha }{4}} \Gamma \left(1+\frac{i \alpha }{2}\right) \Gamma \left(\frac{5-i \alpha }{4} \right), \label{eq:1Rn=0_mu=0_C1}\\
  C_2 &=& -\left(\frac{i}{2}\right)^{\frac{i \alpha }{4}} \Gamma \left(1-\frac{i \alpha }{2}\right) \Gamma \left(\frac{5+i \alpha }{4} \right), \label{eq:1Rn=0_mu=0_C2}
\end{eqnarray}
and several characteristic values of $\alpha$. Here $\Gamma(x)$ stands for the Euler gamma function. At $\mu=0$, the choice of $C_{1,2}$ in the form of Eqs.~\eqref{eq:1Rn=0_mu=0_C1},~\eqref{eq:1Rn=0_mu=0_C2} ensures the convergence of the norm integral; see Appendix~\ref{app:norm}, Eq.~\eqref{eq:R_mu>-3/4}.

\begin{figure}[h]
  \centering
  \includegraphics[width=\columnwidth]{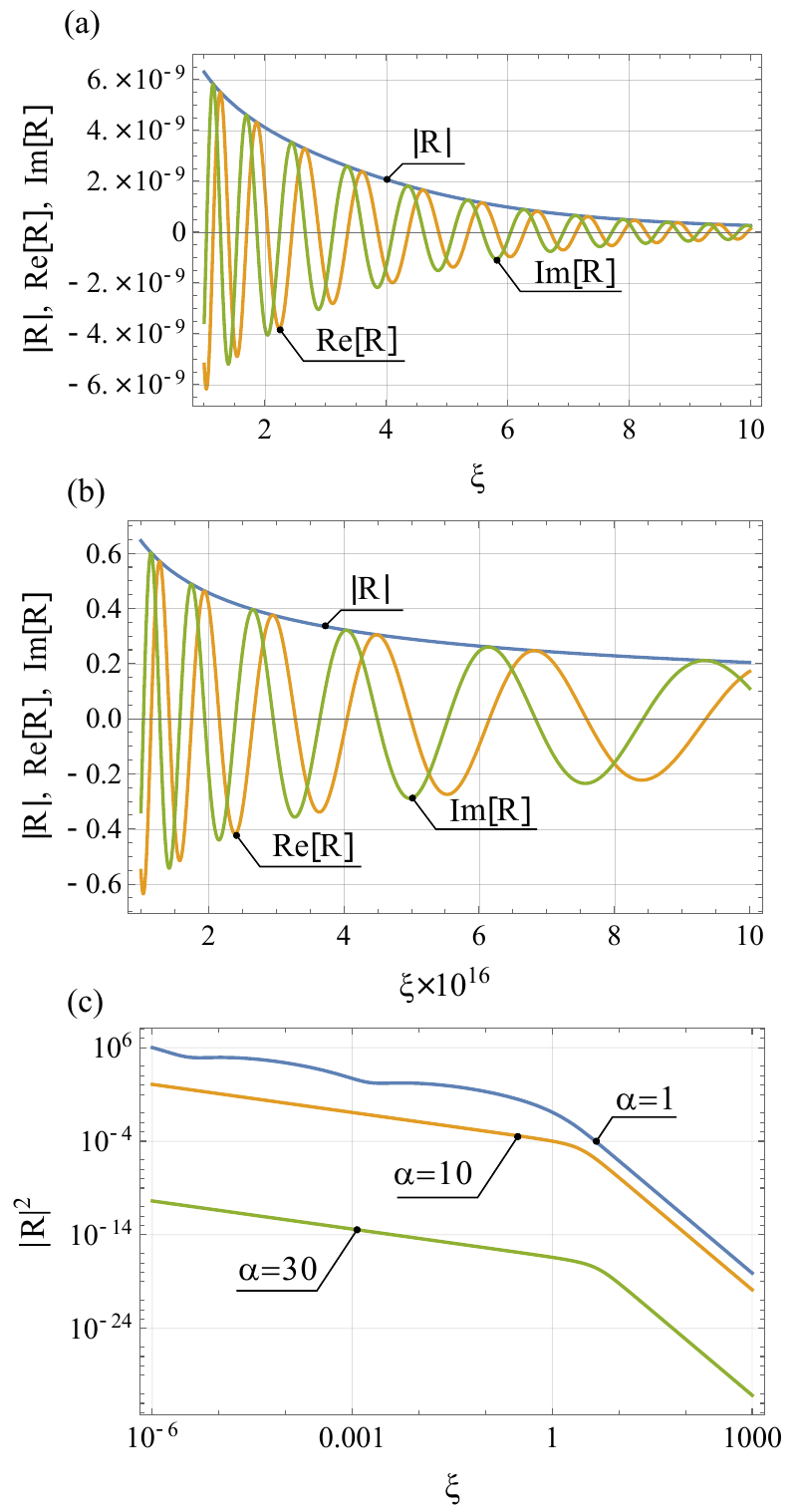}
  \caption{(a) and (b) Modulus, real and imaginary parts of $R(\xi)$ given by Eqs.~\eqref{eq:Sol_R}, in the specific case of Eqs.~\eqref{eq:1Rn=0_mu=0_C1},~\eqref{eq:1Rn=0_mu=0_C2}, $\mu=0$, and $\alpha =30$; {Note the {\it 16 orders of magnitude difference} in the characteristic scales of $\xi$ in (a) and (b).} (c) Log-log plots of $|R(\xi)|^2$ in the case of  Eqs.~~\eqref{eq:1Rn=0_mu=0_C1},~\eqref{eq:1Rn=0_mu=0_C2}, and $\mu=0$, at \mbox{$\alpha=1,\,10,\,30$.} Two power-law asymptotics: $|R|^2 \sim 1/\xi$, at $\xi \rightarrow 0$; and \mbox{$|R|^2 \sim \xi^{-6}$,} at $\xi \rightarrow \infty$ (see Appendix), are well-pronounced.}\label{fig:All}
  %\vspace*{-24pt}
\end{figure}

The corresponding plots are shown in Fig.~\ref{fig:All}. Though the $\xi$-scales in  Figs.~\ref{fig:All}(a) and \ref{fig:All}(b) differ in {\it 16 orders of magnitude}, Re$\,R(\xi)$ and Im$\,R(\xi)$ keep the same self-similar profiles, whose characteristic scale monotonically tends to zero at $\xi \rightarrow 0$. In particular, the phase shift in oscillations of Re$\,R(\xi)$ and Im$\,R(\xi)$ remains fixed at any $\xi$ so that the zeros of one function correspond to the local extrema of the other and vice versa. As a result, the oscillations of the real and imaginary parts of $R(\xi)$ do not affect its modulus, which is a smooth monotonic function of $\xi$. This feature is generic and does not depend on the value of $\alpha$, at least if $\alpha$ is not too small, see Fig.~\ref{fig:All}(c). Notably, such a peculiarity has nothing to do with the self-similarity of Eqs.~\eqref{eq:Self_Psi},~\eqref{eq:R}. The latter affects only the specific choice of $\xi$, not the behavior of $R(\xi)$ at $\xi \rightarrow 0$.
{\section{Particle with zero angular momentum in Coulomb and other subcritical fields} \label{sec:s<2}}

{Here we will discuss the problem peculiarities for a particle with zero angular momentum. In this case, in classical mechanics, a fall to the center occurs at any attractive potential; see Eq.~\eqref{eq:class_cond}. By contrast, the quantum problem does not exhibit collapse if $s$ in Eq.~\eqref{eq:U_s} is less than 2 (we will call such potential subcritical). It is related to the restrictions imposed by the uncertainty relations~\cite{landau2013quantum}. Indeed, let a wave function be localized in a region of radius $r_0$ about the origin. Then, based on the uncertainty relations, its mean energy is estimated as 
\begin{equation}\label{eq:mean_E_s<2}
  E \sim \frac{(\Delta p)^2}{2m} -\frac{\beta}{r_0^s} \sim \frac{\hbar^2}{m r_0^2}-\frac{\beta}{r_0^s}. 
\end{equation}
At $s<2$, this expression is bound from below. It means that the energy spectrum also is bound from below, and the fall to the center is impossible, regardless of the angular momentum value.}

{Nonetheless, at $\ell = 0$, the quantum problem with a subcritical potential has a remarkable peculiarity worth discussing. To highlight it, we consider the stationary version of Schr\"{o}dinger's equation: $\hat{H}\psi = E\psi$. It is well-known that for any fixed $\ell$ and $U(r)$ satisfying the condition $r^2U(r) \rightarrow 0$ at $r\rightarrow 0$, solutions to this equation obey the following asymptotical expression at  $r\rightarrow 0$~\cite{landau2013quantum}:
\begin{equation}\label{eq:psi_at_0}
  \psi \approx \left(\frac{c_1}{r^{\ell+1}} + c_2 r^\ell \right)Y_\ell^m(\theta,\varphi).
\end{equation}
}

{Two terms in delimiters in Eq.~\eqref{eq:psi_at_0} are the first terms of expansions of the two independent solutions to eq.~ $\hat{H}\psi = E\psi$ in powers of small $r$. Usually, $c_1$ is set to zero since the corresponding solution is singular at $r=0$. The remaining solution tends to zero at $r\rightarrow \infty$ only at certain discrete values of $E$. It gives rise to the quantization of the energy spectrum~\cite{landau2013quantum}.}

{However, one of the main points of the present paper is that the fundamentals of quantum mechanics do not require the finiteness of wave functions, provided their norms are finite. On the other hand, at $\ell = 0$, the singularity in Eq.~\eqref{eq:psi_at_0} is square-integrable. Thus, if instead of the finiteness of $\psi$, we require the finiteness of $\|\psi\|$, both terms in Eq.~\eqref{eq:psi_at_0} occur physically meaningful. Then, since the singular at $r=0$ solution tends to zero at $r\rightarrow \infty$, setting $c_2$ to zero results in the required decay at infinity for {\it any} $E<0$. It corresponds to bound states with a {\it continuous\/} spectrum. Note that bound states with a continuous spectrum are a characteristic feature of stationary states for the inverse square potential Eq.~\eqref{eq:U} admitting quantum collapse~\cite{book:Morse_Feshbach}.  }

{To be more specific, we consider the Coulomb field with $U(r) = -\beta/r$ as the most essential example. In this case, measuring mass, length, time, and energy in units of $m,\;\frac{\hbar^2}{m\beta},\;\frac{\hbar^3}{m\beta^2}$, and $\frac{m\beta^2}{\hbar^2}$, respectively, at $\ell=0$, we reduce the equation $\hat{H}\psi = E\psi$ to the following form~\cite{landau2013quantum}: 
\begin{equation}\label{eq:psi_Coulomb}
  \psi'' +\frac{2}{\rho}\psi' +\left(\frac{n}{\rho}-\frac{1}{4}\right)\psi = 0.
\end{equation}
Here $n=1/\sqrt{-2E},\;\rho = 2r/n,\;E<0$ means $d/d\rho$. Note also that, at $\ell = 0$, $\psi$ does not depend on the angular variables.}

{According to those mentioned above, a regular at $\rho = 0$ and $\rho \rightarrow \infty$ solution to Eq.~\eqref{eq:psi_Coulomb} exists only at integer $n\geq 0$, which gives rise to the well-known discrete spectrum of a Hydrogen atom. However, here we are interested in the other, singular at $r=0$ solution. It reads as follows:
\begin{equation}\label{eq:psi_Coulomb_singular}
 \psi(\rho) = c \exp{\left(-\frac{\rho}{2}\right)} U(1-n,2,\rho),
\end{equation}
where $U(a,b,z)$ stands for the Tricomi confluent hypergeometric function (do not confuse it with the potential $U(r)$).}

{Expression \eqref{eq:psi_Coulomb_singular} has the following asymptotics:
\begin{eqnarray}
   & & \psi(\rho) \approx c_1 \rho^{n-1}\exp{\left(-\frac{\rho}{2}\right)}, \;\;\mbox{at} \; \rho \rightarrow, \infty \label{eq:psi_infty}\\
   & & \psi(\rho) \approx -\frac{c_1}{n\rho\Gamma(-n)}, \;\;\mbox{at} \; \rho \rightarrow 0 \label{eq:psi_Coulomb_at_0}.
\end{eqnarray}
Interestingly, Eq.~\eqref{eq:psi_Coulomb_at_0} is not valid for integer $n\geq 0$ (i.e. precisely for the values of $n$ corresponding to the discrete spectrum; see above) since $|\Gamma(-n)|$ in the denominator terns to infinity. Asymptotic Eqs.~\eqref{eq:psi_infty},~\eqref{eq:psi_Coulomb_at_0} agree with the general discussion at the beginning of this section and provide square-integrability of $\psi(\rho)$ since $U(1-n,2,\rho)$ does not have singularities at $0<\rho<\infty$. That is, $\|\psi\|$ remains finite.} 

{Thus, the obtained finite-norm exact solution Eq.~\eqref{eq:psi_Coulomb_singular} to Schr\"{o}dinger's equation has a continuous, unbound from below spectrum of negative energy values and corresponds to bound states in a Coulomb field. This result directly contradicts the above conclusion that quantum collapse in subcritical potentials is impossible, to say nothing about the extremely well-known experimental spectrum of a hydrogen atom. How to resolve the contradiction?   }

{The point is that solution Eq.~\eqref{eq:psi_Coulomb_singular} does not satisfy the uncertainty relations: the integral determining $\langle p_r\rangle$ diverges at the lower limit, and $\Delta p_r$ does not exist. Then, the uncertainty relations cannot be applied. It means, this 'solution' does not correspond to any actually realizable quantum state and should be dropped as non-physical. This example indicates that, though a wave function indeed may be singular, to be physically meaningful, in addition to the finiteness of its
norm, one must request that, in the corresponding quantum state, mean values of observables are also finite, and the uncertainty relations hold.}
\vspace*{12pt}

\section{Conclusions\label{sec:concl}}

Thus, we have studied and compared non-regularized classical and quantum collapses {(escapes). We constructively have proven that close to the completion of the collapse (the beginning of the escape), the general solution to the classical problem is transformed into the self-similar one, i.e. the latter is an attractor for any other solutions. We also have shown that the classical collapse is continuously transformed into the escape and vice versa, i.e. the end of either marks the beginning of the other. Calculating the radiative losses for a specific example of the collapse corresponding to the fall of an electron with zero angular momentum to a neutral atom (molecule) with a finite dipole moment, we have shown that long before the impact of the radiative losses
on the collapse dynamics becomes noticeable, the classical description must be replaced by the quantum one.}

Then, we have proven the existence of quantum collapse and escape by obtaining the family of exact solutions to Schr\"{o}dinger's equation, describing the phenomena. By simple arguments of dimensional analysis  {supplemented by the matching of the general quantum and classical cases through the quasi-classical approximation}, we have shown that the obtained self-similar solutions to the quantum problem also are attractors for a much broader class of non-self-similar ones.}

Since, for the obtained exact solutions to Schr\"{o}dinger's equation, the norm of the wave function is time-dependent, we have generalized to this case the conventional rule to calculate the mean values of operators, derived the corresponding continuity equation, and discussed its properties. {We also have unveiled the crucial role of the uncertainty relations in a selection of correct rules generalizing conventional laws of quantum mechanics to unconventional cases and in the separation of physically meaningful and meaningless solutions to Schr\"{o}dinger's equation.}    

{We have revealed a striking similarity between the classical and quantum collapses. Presumably, this fact is related to the problem symmetry, insensitive to its classical or quantum nature.} 

{Note that two- and one-dimensional versions of quantum collapse are also meaningful~\cite{parisi1973anomalous,van1978bound,PhysRevD.68.125013,AvilaAoki2009,sakaguchi2011suppression2,nisoli2014attractive}. Since the spatial dimension does not affect the scaling properties of Schr\"{o}dinger's equation required for self-similarity, the approach developed in the present paper may be straightforwardly applied to these problems too. In these cases, the self-similar variable $\xi$ remains the same, while the governing equation~\eqref{eq:R} becomes different. Detailed discussions of these issues lie beyond the scope of our analysis.} 

{Thus, the self-similar solutions introduced here are a convenient and powerful tool to investigate various dynamical effects in classical and quantum mechanics. Hopefully, the presented study sheds new light on the fundamentals of quantum mechanics and provides a better understanding of its basic principles.}

\begin{acknowledgments}
The author is grateful to \framebox[1.1\width]{L.P. Pitaevskii}, S.A.~Sigov, {N.V.~Brilliantov,} A.M.~Kamchatnov, E.A.~Kuznetsov, B.A.~Malomed, B.Y.~Rubinstein, Y.A.~Stepanyants, and, last but not least, G.E.~Volovik for the fruitful discussions of this work. The work was supported by the Russian Foundation for Basic Research [Projects No. 20-02-00086] for the analytical study, the Russian Science Foundation [Project No. 21-12-00151] for the symbolic computer calculations, {and the Ministry of Science and Higher Education of the Russian Federation [Agreement No. 075-15-2022-1150] for numerics. The investigation of peculiarities of Non-Hermitian Hamiltonians is supported by the Russian Science Foundation [Project No. 23-72-00037].}
\end{acknowledgments}

\appendix
\renewcommand\theequation{\Alph{section}.\arabic{equation}}
\section{Self-Similar version of Schr\"{o}dinger's equation\label{app:self-sim}}

Here we find the conditions for the reduction of Schr\"{o}dinger's equation to a self-similar form. One of the most general ansatzes for the wave function with a given $\ell$ is as follows:
\begin{equation}\label{eq:Phi(t)R(xi)}
 \Psi = \Phi(-\chi t ) R(\xi )Y_{\ell }^m(\theta ,\varphi),
\end{equation}
where $\xi =r/(-\chi t)^{\nu } $, while $\Phi$ and $R$ are yet unknown functions. For this type of $\Psi(r,t)$ we have \mbox{$\partial^n \Psi/\partial r^n = (-\chi t)^{-n\nu }\partial^n \Psi/\partial \xi^n$}. Regarding ${\partial \Psi}/{\partial t}$, it is given by the following expression:
\begin{equation}\label{eq:d/dt}
  \frac{\partial \Psi}{\partial t} = \frac{\nu  \xi  \chi  \Phi (z) R'(\xi )}{z}-\chi  R(\xi ) \Phi '(z).
\end{equation}
Here $z \equiv -\chi t$.
Our goal is to reduce Schr\"{o}dinger's equation to an equation depending solely on $\xi$ and independent of $t$ explicitly. The necessary condition for that is that $\Phi(z)/z$ and $\Phi '(z)$ on the l.h.s of Schr\"{o}dinger's equation have the same dependence on $t$ so that it makes a common for them factor, which may be canceled with the same factor on the r.h.s. It means that
\begin{equation}\label{eq:eq_for_Phi}
  \frac{d\Phi}{dz}=\mu \frac{\Phi}{z},
\end{equation}
where $\mu$ is an arbitrary constant. Integrating this equation, we readily obtain
\begin{equation}\label{eq:sol_fo_Phi}
  \Phi = const\cdot z^\mu
\end{equation}
Substituting Eq.~\eqref{eq:Phi(t)R(xi)} with this form of $\Phi(z)$ into Eq.~\eqref{eq:Schr} with $U(r)$ given by Eq.~\eqref{eq:U_s} we arrive at the following equation:

\begin{eqnarray}
   & & z^{1-2 \nu } R''(\xi )+ 2\left[\frac{ i m \nu  \xi  \chi}{\hbar }+\frac{ z^{1-2 \nu }}{\xi }\right] R'(\xi )\label{eq:SE at any_nu} \\
  &+& \left[\frac{2 \beta  m  z^{1-\nu  s}}{\hbar ^2 \xi^s}-\frac{2 i \mu  m \chi }{\hbar }-\frac{\ell  (\ell +1) z^{1-2 \nu }}{\xi ^2}\right]R(\xi)=0.\nonumber
\end{eqnarray}
It is seen straightforwardly that Eq.~\eqref{eq:SE at any_nu} does not depend on $z$ if and only if $\nu = 1/2$ and $s=1/\nu$=2. %, cf. Eq.~\eqref{eq:Self-sim_cond_Phi}.
Regarding $\chi$, the choice $\chi = \hbar/m$ is just a matter of convenience to turn the corresponding coefficient to unity. Then, Eq.~\eqref{eq:SE at any_nu} is transformed into Eq.~\eqref{eq:R}.

\section{Norm Convergence\label{app:norm}}

Here we discuss the convergence of $\int_{0}^{\infty}|R(\xi)|^2\xi^2d\xi$, where $R(\xi)$ is given by Eq.~\eqref{eq:Sol_R}. Since $_1\!F_1(a;b;z)$ is an analytic function of $z$ on the whole complex plane~\cite{janke1960tafeln} only the convergence at the lower and upper limits should be examined.

The lower limit case is simple. Taking into account that  $_1\!F_1(a;b;z) = 1 + O(z)$  at $z \rightarrow 0$ \cite{janke1960tafeln}, we readily obtain that in proximity of $\xi=0$ the most singular terms in the solution give rise to the expression
\begin{equation}\label{eq:R_at_0}
  |R(\xi)| \approx \frac{1}{\sqrt{\xi}}\left|C_1 \xi^{-\frac{i \alpha }{2}} + C_2 \xi^{\frac{i \alpha }{2}}\right| \leq \frac{|C_1|+|C_2|}{\sqrt{\xi}}.
\end{equation}

This means, the singularity of $|R(\xi)|^2\xi^2$ at $\xi=0$ is integrable. Remarkably, estimate \eqref{eq:R_at_0} does not depend on $\mu$. Thus, the integral $\int_{0}^{\infty}|R(\xi)|^2\xi^2d\xi$ converges at the lower limit at any $\mu$.

The case $\xi \rightarrow \infty$ is more tricky. The asymptotic expansion of $_1\!F_1(a;b;z)$ at $|z| \rightarrow \infty$ reads~\cite{janke1960tafeln}
\begin{eqnarray}
   _1\!F_1(a;b;z) &=& \frac{\Gamma(b)}{\Gamma(b-a)}(-z)^{-a}G(a,a-b+1,-z)\nonumber \\
   & & + \frac{\Gamma(b)}{\Gamma(a)}e^z z^{a-b}G(b-a,1-a,z), \label{eq:1F1_assympt}
\end{eqnarray}
where %$\Gamma(x)$ stands for the Euler gamma function and
\begin{equation}\label{eq:G}
  G(a,b,z)\!=\!1+\frac{ab}{1!z}+\frac{a(a+1)b(b+1)}{2!z^2}+\ldots\!=\!\sum_{n=0}^{\infty}\frac{(a)_n(b)_n}{n!z^n}.
\end{equation}
In Eq.~\eqref{eq:G} $(x)_n$ designates the rising factorial (the Pochhammer function) defined as follows:
\begin{eqnarray}
  (x)_0 &=& 1,\nonumber \\
  (x)_n &=& x(x+1)(x+2)\cdots(x+n-1)\label{eq:rising!}\\
  &=&\prod_{k=0}^{n-1}(x+k) = \frac{\Gamma(x+n)}{\Gamma(x)}.\nonumber
\end{eqnarray}

Then, according to Eqs.~\eqref{eq:1F1_assympt},~\eqref{eq:G}, the asymptotic expansion for $_1\!F_1(a;b;z)$ has two groups of terms originated in the expressions proportional to $(-z)^{-a}G$ and $e^z z^{a-b}G$, respectively. According to Eq.~\eqref{eq:G}, to select the most singular terms in every group at $z \rightarrow \infty$, we have to replace $G$ by 1. Then, Eq.~\eqref{eq:Sol_R} results in the following asymptotic expression for $R(\xi)\approx\, _1R^{(0)}(\xi)+\,_2R^{(0)}(\xi)$, where
\begin{eqnarray}
   &_1R^{(0)}(\xi)&=\,_1C^{(0)} \xi ^{2 \mu }\label{eq:1R0}\\
   & &\!\!\!\!\!\!\!\!\!\!\!\!\!\!\!\!\!\!\times\left[\frac{\left(\frac{i}{2}\right)^{\frac{i \alpha }{4}} \Gamma \left(1-\frac{i \alpha }{2}\right)}{\Gamma \left(\frac{5-i \alpha }{4}+\mu\right)}{C_1}+\frac{\left(\frac{i}{2}\right)^{-\frac{i \alpha}{4}} \Gamma \left(1+\frac{i \alpha }{2}\right)}{\Gamma \left( \frac{5+i \alpha }{4}+\mu\right)}{C_2}\right],\nonumber\\
%\end{eqnarray}
%\begin{eqnarray}
   &_2R^{(0)}(\xi)&=\,_2C^{(0)}e^{-\frac{i \xi ^2}{2}} \xi ^{-3-2 \mu}\label{eq:2R0}\\
   & &\!\!\!\!\!\!\!\!\!\!\!\!\!\!\!\!\!\!\times\left[\frac{(-\frac{i}{2})^{\frac{i \alpha}{4}}  \Gamma \left(1-\frac{i \alpha }{2}\right)}{\Gamma \left(-\frac{1+i \alpha }{4}-\mu\right)}{C_1}
  +\frac{\left(-\frac{i}{2}\right)^{-\frac{i \alpha}{4}} \Gamma \left(1+\frac{i \alpha }{2}\right)}{\Gamma \left(-\frac{1-i \alpha }{4}-\mu\right)}{C_2}\right],\nonumber
\end{eqnarray}
Here $_{1,2}C^{(0)}$ are constants. {The cumbersome explicit expressions for them are obtained} upon the step-by-step implementation of the above procedure. We do not need these expressions for the further analysis.

With the proper choice of the ratio $C_1/C_2$ we turn to zero either $_1R^{(0)}(\xi)$ or $_2R^{(0)}(\xi)$. If instead of the leading terms solely, we employ the entire infinite series Eq.~\eqref{eq:G}, the expression for $R(\xi)$ becomes the following:
\begin{equation}\label{eq:R_series}
  R(\xi)=\sum_{n=0}^{\infty} [_1\!R^{(n)}(\xi)+\,_2R^{(n)}(\xi)].
\end{equation}
A remarkable thing, however, is that the expressions for $_{1,2}R^{(n)}(\xi)$ preserve the same structure as that in Eqs.~\eqref{eq:1R0}, \eqref{eq:2R0}. The only difference is in the change of the prefactor: $_1C^{(0)} \xi ^{2 \mu } \rightarrow \, _1C^{(n)} \xi ^{2 (\mu-n) }$ for $_{1}R^{(n)}(\xi)$ and $_2C^{(0)}e^{-\frac{i \xi ^2}{2}} \xi ^{-3-2 \mu} \rightarrow \, _2C^{(n)}e^{-\frac{i \xi ^2}{2}} \xi ^{-3-2 (\mu+n)}$ for $_{2}R^{(n)}(\xi)$; the expressions in the square delimiters in  Eqs.~\eqref{eq:1R0}, \eqref{eq:2R0} do not depend on $n$ and hence remain the same at any $n$. It means, that the value of the ratio $C_2/C_1$ which turns $_j\!R^{(0)}$ to zero (here $j=1,2$) simultaneously turns to zero all $_j\!R^{(n)}$ with the same value of $j$ and any value of $n$, i.e. the entire infinite series $\sum_{n=0}^{\infty}{_j\!R^{(n)}(\xi)}$ vanishes.

Thus, at
\begin{equation}\label{eq:2Rn=0}
  \frac{C_2}{C_1}=-\frac{\left(-\frac{i}{2}\right)^{\frac{i \alpha }{2}} \Gamma \left(1-\frac{i \alpha }{2}\right) \Gamma \left(-\frac{1-i \alpha }{4}-\mu \right)}{\Gamma \left(1+\frac{i \alpha }{2}\right) \Gamma \left(-\frac{1+i \alpha }{4}-\mu \right)},
\end{equation}
$_{2}R^{(n)}=0$ at any $n$, and $ R(\xi)=\sum_{n=0}^{\infty} \,_1R^{(n)}(\xi)$.

At
\begin{equation}\label{eq:1Rn=0}
  \frac{C_2}{C_1}=-\frac{\left(\frac{i}{2}\right)^{\frac{i \alpha }{2}} \Gamma \left(1-\frac{i \alpha }{2}\right) \Gamma \left(\frac{5+i \alpha }{4}+\mu \right)}{\Gamma \left(1+\frac{i \alpha }{2}\right) \Gamma \left(\frac{5-i \alpha }{4}+\mu \right)},
\end{equation}
all $_{1}R^{(n)}$ vanish, and $ R(\xi)=\sum_{n=0}^{\infty} \,_2R^{(n)}(\xi)$.

In both cases, the most singular term is the one with $n=0$. Let us inspect these terms' contribution to the norm's integral for the obtained wave function. In case Eq.~\eqref{eq:2Rn=0}, $|R(\xi)|^2 \sim \xi^{4\mu'}$. Then, at $\xi \rightarrow \infty$, the integral $\int |R(\xi)|^2\xi^2d\xi \sim \xi^{4\mu' +3}$. Its convergence requires \mbox{$\mu' <-3/4$.}

Similarly, in case Eq.~\eqref{eq:1Rn=0}, $|R(\xi)|^2 \sim \xi^{-6-4\mu'}$. Then, the norm integral converges at the upper limit, provided  \mbox{$\mu' > -3/4$}.

At $\mu'=-3/4$ the integral $\int_{0}^{\infty}|R(\xi)|^2\xi^2d\xi$ diverges at the upper limit as $\ln\xi$ owing to the contribution of $|_{1,2}R^{(0)}(\xi)|^2$. Since the sole value of $\mu'$ when the norm of the obtained wave function does not depend on time is $\mu'=-3/4$, only time-dependent norms are physically meaningful for the given wave function.

It is convenient to present the explicit form of the obtained solutions admitting the normalization. It is as follows:

At $\mu'<-3/4$
\begin{eqnarray}
   R(\xi) & = & \frac{C}{\sqrt{\xi}}\left[\left(-\frac{i\xi^2}{2}\right)^{-\frac{i\alpha}{4}}\Gamma\left(1+\frac{i\alpha}{2}\right)
   \Gamma\left(-\frac{1+i\alpha}{4} -\mu\right)\right.\nonumber \\
  &\times & \left.\!\!\!\! _1F_1\left(-\frac{1+i \alpha }{4}-\mu ;1-\frac{i \alpha }{2};-\frac{i\xi ^2}{2} \right)\right.\label{eq:R_mu<-3/4} \\
  &-&\!\! \left(-\frac{i\xi^2}{2}\right)^{\frac{i\alpha}{4}}\Gamma\left(1-\frac{i\alpha}{2}\right)
   \Gamma\left(-\frac{1-i\alpha}{4} -\mu\right)\nonumber \\
   &\times & \left.\!\!\!\! _1F_1\left(-\frac{1-i \alpha }{4}-\mu ;1+\frac{i \alpha }{2};-\frac{i\xi ^2}{2} \right)\right]. \nonumber
\end{eqnarray}
At $\mu'>-3/4$
\begin{eqnarray}
   R(\xi) & = & \frac{C}{\sqrt{\xi}}\left[\left(\frac{i\xi^2}{2}\right)^{-\frac{i\alpha}{4}}\Gamma \left(1+\frac{i \alpha }{2}\right) \Gamma \left(\frac{5-i \alpha }{4}+\mu\right)\right.\nonumber \\
  &\times & \left.\!\!\!\! _1F_1\left(-\frac{1+i \alpha }{4}-\mu ;1-\frac{i \alpha }{2};-\frac{i\xi ^2}{2} \right)\right. \label{eq:R_mu>-3/4}\\
  &-&\!\! \left(\frac{i\xi^2}{2}\right)^{\frac{i\alpha}{4}} \Gamma \left(1-\frac{i \alpha }{2}\right) \Gamma \left(\frac{5+i \alpha }{4}+\mu\right)\nonumber \\
   &\times & \left.\!\!\!\!  _1F_1\left(-\frac{1-i \alpha }{4}-\mu ;1+\frac{i \alpha }{2};-\frac{i\xi ^2}{2} \right)\right]. \nonumber
\end{eqnarray}
For the sake of symmetry, we have rescaled the constant of integration in Eqs.~\eqref{eq:1R0},~\eqref{eq:2R0} so that for Eq.~\eqref{eq:R_mu<-3/4}
\begin{equation}\label{eq:C_mu<-3/4}
  C_1=\left(-\frac{i}{2}\right)^{-\frac{i \alpha }{4}} \Gamma \left(1+\frac{i \alpha }{2}\right) \Gamma \left(-\frac{1+i \alpha }{4}-\mu \right)C
\end{equation}
while for Eq.~\eqref{eq:R_mu>-3/4}
\begin{equation}\label{eq:C_mu>-3/4}
  C_1=\left(\frac{i}{2}\right)^{-\frac{i \alpha }{4}}  \Gamma \left(1+\frac{i \alpha }{2}\right) \Gamma \left(\frac{5-i \alpha }{4}+\mu \right)C
\end{equation}

Note that the expressions in square delimiters on the r.h.s.' of both Eq.~\eqref{eq:R_mu<-3/4} and Eq.~\eqref{eq:R_mu>-3/4} are the differences of the two terms, where the second term is obtained from the first by the formal transformation $\alpha \rightarrow -\alpha$. At $\alpha = 0$ they are identical, and  the  r.h.s.' of Eqs.~\eqref{eq:R_mu<-3/4},~\eqref{eq:R_mu>-3/4}  vanish. Since the necessary collapse condition reads $\alpha \equiv \sqrt{4\gamma - 1} >0$, see Eq.~\eqref{eq:alpha}, it means that the obtained solutions smoothly vanish at the continuous transformation of the potential from the collapsing to the non-collapsing type at $\alpha \rightarrow 0$.
\mbox{}\\

{\section{Uncertainty relations\label{app:uncert}}}
{For the reader's convenience, we reproduce here the main points of the derivation of the uncertainty relations presented in~\cite{landau2013quantum}, to show that it indeed does not require the Hamiltonian self-adjointness. First of all, we recall that in the coordinate representation, the momentum operator reads
\begin{equation}\label{eq:p_hat}
  \hat{\mathbf{p}} =-i\hbar\frac{\partial}{\partial\mathbf{r}},
\end{equation}}
\mbox{}

{Then, we suppose that a particle with the mean value of the momentum $\mathbf{p}_0$ is localized in a finite region of space with the sizes $\Delta x$, $\Delta y$, and $\Delta z$. In this case, its wave function has the form $\psi = u(\mathbf{r})\exp[(i\mathbf{rp}_0/\hbar)]$, where $u(\mathbf{r})$ sharply decays beyond the localization region.}

{Next, the eigenfunctions of the operator $\hat{\mathbf{p}}$ are plane waves so that the corresponding eigenfunction expansion of a wave function  $\psi$ is just a Fourier integral. The coefficients in this integral $\psi_{\mathbf{p}}$ are the Fourier transforms \mbox{of $\psi$:}
\begin{equation}\label{eq:psi_p}
  \psi_{\mathbf{p}} = \int u(\mathbf{r})\exp[i\mathbf{r}(\mathbf{p}_0-\mathbf{p})/\hbar]d^3\mathbf{r}
\end{equation}
Since the mentioned sharp decay of $u(\mathbf{r})$ beyond the localization region, this region makes an overwhelming contribution to the integral on the r.h.s. of Eq.~\eqref{eq:psi_p}. On the other hand, if we consider $\psi_{\mathbf{p}}$ as a function of $\Delta p_{x,y,z}=|p_{0x,y,z}-p_{x,y,z}|$, we will see its rapid decay as soon as $\Delta p_{x,y,z}$ exceed the values
\begin{equation}\label{eq:uncert}
  {\Delta p_x\Delta x \sim \hbar},\; {\Delta p_y\Delta y \sim \hbar}, \; {\Delta p_z\Delta z \sim \hbar}.
\end{equation}
The decay is related to the rapid oscillations of $\exp[i\mathbf{r}(\mathbf{p}_0-\mathbf{p})/\hbar]$ in the area of the main contribution to the integral, if $\Delta p_{x,y,z}$ occurs beyond the specified bounds.}

{To complete the derivation, we have to recall that the probability density to find a given value of $\mathbf{p}$ is proportional to $|\psi_{\mathbf{p}}|^2$. Thus, the probability for $\Delta p_{x,y,z}$ to have a value beyond the bounds Eq.~\eqref{eq:uncert} is negligibly small. }

\bibliography{FallRef_new_F}
\end{document}